\documentclass[aps,pre,groupedaddress,twocolumn]{revtex4}
\pdfoutput=1

\usepackage{hyperref}
\usepackage{subfigure}
\usepackage{color}
\usepackage{graphicx}
\usepackage[]{natbib}

\begin{document}
\bibliographystyle{unsrtnat}
\expandafter\ifx\csname urlprefix\endcsname\relax\def\urlprefix{URL }\fi

\DeclareGraphicsExtensions{.pdf, .jpg}

\title{\large
   Matter waves in the Talbot-Lau interferometry.
    }

\large
\author{Valeriy I. Sbitnev}
\email{valery.sbitnev@gmail.com}
\affiliation{B.P.Konstantinov St.-Petersburg Nuclear Physics Institute, Russ. Ac. Sci.,
     Gatchina, Leningrad district, 188350, Russia.}


\date{\today}

\begin{abstract}
 Paths of particles, emitted from distributed sources and passing out through slits of two gratings, $G_{0}$ and $G_{1}$, up to detectors, have been computed in details by the path integral method.
 The slits are represented by Gaussian functions that simulate fuzzy edges of the slits.
 Waves of matter be computed by this method show perfect interference patterns both between the gratings and behind the second grating. Coherent and noncoherent the distributed particle sources reproducing the interference patterns are discussed in details. Paraxial approximation stems from the wave function when removing the distributed sources onto infinity.
 The more hard-edged slits of the grating $G_{1}$ are examined by simulating those slits  by a superposition of more hard the Gaussian functions.

 As for the particles here we consider fullerene molecules.
 De Broglie wavelength of the molecules is adopted equal to 5 pm.


\end{abstract}

\maketitle

\large

\section{\label{sec:level1}Introduction}

 Interferometry with matter waves, where particles are presented by heavy molecules, such as fullerene molecules, attracts the last years a great interest of scientific community; see review of \citet*{CroninEtAl2009}, and rich set of references ibid.
 Interferometric experiments help to disclose the very basic principles of quantum physics with systems of rather large size and
 complexity~\citep{ NowakEtAl1997, HornbergerEtAl2004, BrezgerEtAl2003, ZeilingerEtAl2002, HornbergerEtAl2003, ArndtEtAl2005, ZeilengerEtAl2003}.

\begin{figure}[htb!]
  \centering
  \begin{picture}(200,60)(-50,10)
      \includegraphics[scale=0.5]{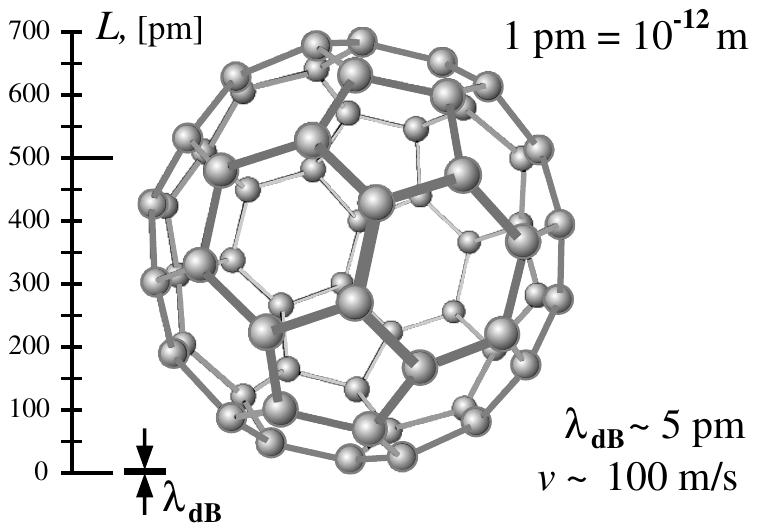}
  \end{picture}
  \caption{
  The fullerene molecule C$_{_{60}}$ consists of 60 carbon atoms.
  From the left, a characteristic size of the molecule together with its characteristic de Broglie wavelength are shown.
  }
  \label{fig=1}
\end{figure}

 Heavy molecules, having masses about 100 amu and more, are particles showing under ordinary circumstances almost classical behavior.
 Indeed, diameter of the fullerene molecule C$_{60}$, see Fig.~\ref{fig=1}, is about 0.7~nm~\citep{YanovLeszczynski2004}, but de Broglie wavelength is $\sim5$~pm~\citep{HornbergerEtAl2003, JuffmannEtAl2010} (velocity of the molecule is about 100~m/s).
 The wavelength is shorter of the diameter by a factor of 100
 and much smaller than distances between atoms in a solid body, whereas a size of the molecule exceeds these distances.
 Interference fringes emergent at scattering such heavy particles on a grating bear an important information
 about the wave-particle duality of these particles.
 Therefore, in order to reveal wave-particle duality for the heavy particles, interferometers should possess by heightened requirements~\citep{ArndtEtAl2005}.

 By adopting the wave-particle duality formalism relative to the heavy particles, one may then deal with flows of such particles as waves
 incident on the slit grating~\citep{Hornberger2004, Hornberger2005, Hornberger2008, Hornberger2009}.
 Methods based on computing the Fourier images of general two-dimensional periodic objects
 in light and electron optics~\citep{CowleyMoodie1957a, CowleyMoodie1957b, CowleyMoodie1957c, CowleyMoodie1960}
 can be proposed as good algorithms for finding interference patterns induced by the matter waves.
 Next, a method adopted for the analytical description of diffraction uses the Kirchhoff-Fresnel integral~\citep{HornbergerEtAl2003}.
 The integral represents spherical waves emitted from a radiation source.
 Because the integral is taken only over the slit, the slit appears to act like an effective source of radiation, in accordance with the Huygens principle~\citep{Schuocker1998}.
 This principle states that wave fronts passing through slits act as effective sources, reradiating spherical waves.

 It is not intuitively obvious that the particles passing through the slits should be deflected to recreate spherical wave patterns.
 Paradox is that how does the particle identify positions of all slits?~\citet*{ZeilengerEtAl2003} have written "The wave-particle duality states that the description of one and the same physical object suggests the local particle picture in the source and on the screen, but a wave model for the unobserved propagation of the object."
 While the particles go through the slits one by one, a single particle always gives a single click at a detector.
 Between the source and the detector history of the particle remains unknown.
 We may consider all paths of point-like particles going through the slits and converging into the detector.
 This task is solved by the path integral method~\citep{Feynman1948, FeynmanHibbs:1965}.
 The converging paths at the detector give rise to superposition of different histories, which produce an interference phenomenon.

\begin{figure}[htb!]
  \centering
  \begin{picture}(200,100)(-10,15)
      \includegraphics[scale=0.35]{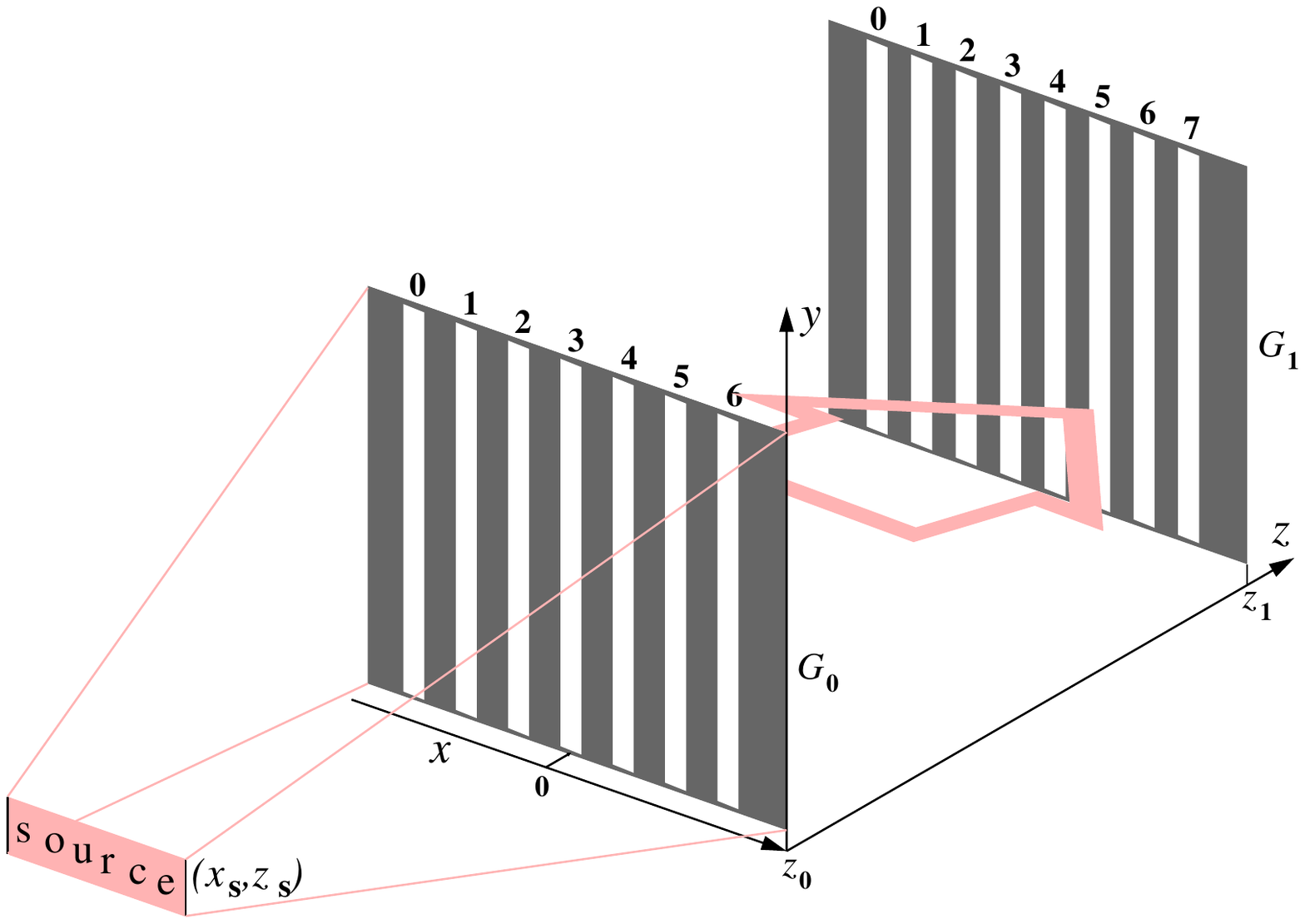}
  \end{picture}
  \caption{
  The Talbot-Lau interferometer scheme:
  two gratings, $G_{0}$ and $G_{1}$ are situated in consecutive order along a particle beam emitted from a distributed source.
  }
  \label{fig=2}
\end{figure}
 Here we will consider emergence of the interference patterns at passing fullerene-like particles through two gratings, $G_{0}$ and $G_{1}$,
 placed in consecutive order along a particle beam, see Fig.~\ref{fig=2}.
 It is a typical scheme of the Talbot-Lau interferometer~\citep{JahnsLohmann1979, McMorranCronin2009}.
 For finding the interference pattern we will compute all paths going from
 the source to the detector screen, i.e., we will compute the path integrals.
 Emergence of the interference patterns is considered in the near-field -- distance between the grating is half of the Talbot length,
 $z_{\rm T}/2=d^{\,2}/\lambda_{\rm dB}=z_{1}-z_{0}$. Here $\lambda_{\rm dB}$ is de Broglie wavelength and $d$ is a distance between the slits.

 Coherence properties of the beam are indispensable conditions for manifestation of the wave-particle duality and, as an effect, emergence of interference fringes. These properties depend on the source, collimation slits, and other extra conditions~\citep{ZeilengerEtAl2003}.
 Ones distinguish spatial coherence and spectral coherence. The first is conditioned by finite width of the source that can exceed de Broglie wavelength.
 And the second can deteriorate because of generating the particles with different escaping velocities.
 Here we will try to reproduce the both coherence properties.
 We will consider partially coherent sources,
 referred to as the generalized Gaussian Schell-model (GSM) sources~\citep{MandelWolf1995, GburWolf2001, McMorranCronin2008a, McMorranCronin2008b}.
 Such sources reproduce planar GSM beams.
 These beams are characterized by two signs,
 that are an intensity distribution across the beams and a spectral degree of the coherence.
 The both are represented by the Gaussian distributions with their own dispersion constants $\sigma_{I}$ and $\sigma_{g}$. With suitable choices of $\sigma_{I}$ and $\sigma_{g}$, such a source will simulate a beam, called GSM beam~\citep{MandelWolf1995, GburWolf2001}.

 Observe, however, that the particle beam represents, in accordance with the particle-wave duality, a matter wave,
 that spreads through the grating device. It should be noted, that the wave consists of two parts, amplitude and phase ones multiplied each other,
 $|\Psi\rangle=\sqrt{\rho}\,\exp\{{\bf i}S/\hbar\}$. Only on the detector we register the intensity proportional to $\rho=\langle\Psi|\Psi\rangle$.
 We will apply the Gaussian Schell-model to the wave function as a simple Gaussian form-factor loaded by either of two dispersion constants $\sigma_{I}$ and $\sigma_{g}$.

 The article is organized as follows.
 In Sec.~\ref{sec:level2} we compute passing the particles through the two gratings, $G_{0}$ and $G_{1}$, being emitted from distributed sources.
 Exact formulas describing interference effect emergent both between the gratings and behind the second grating are represented in this section.
 Sec.~\ref{sec:level3} represents calculations of interference patterns emergent at scattering fullerene-like particles on the gratings
 with finite amount of slits.
 Here we consider an effect of spatial coherence to interference patterns.
 In Sec~.\ref{sec:level4} we remove the sources to infinity and obtain the formulas for interference in the paraxial approximation.
 Here we consider an effect of spectral coherence.
 Sec.~\ref{sec:level5} discusses interference from the gratings with more hard-edged slits.
 Sec.~\ref{sec:level6} is concluding.

\section{\label{sec:level2}Path integral: particles passing through slits in two screens}

\begin{figure}[htb!]
  \centering
  \begin{picture}(200,120)(-20,15)
      \includegraphics[scale=0.5]{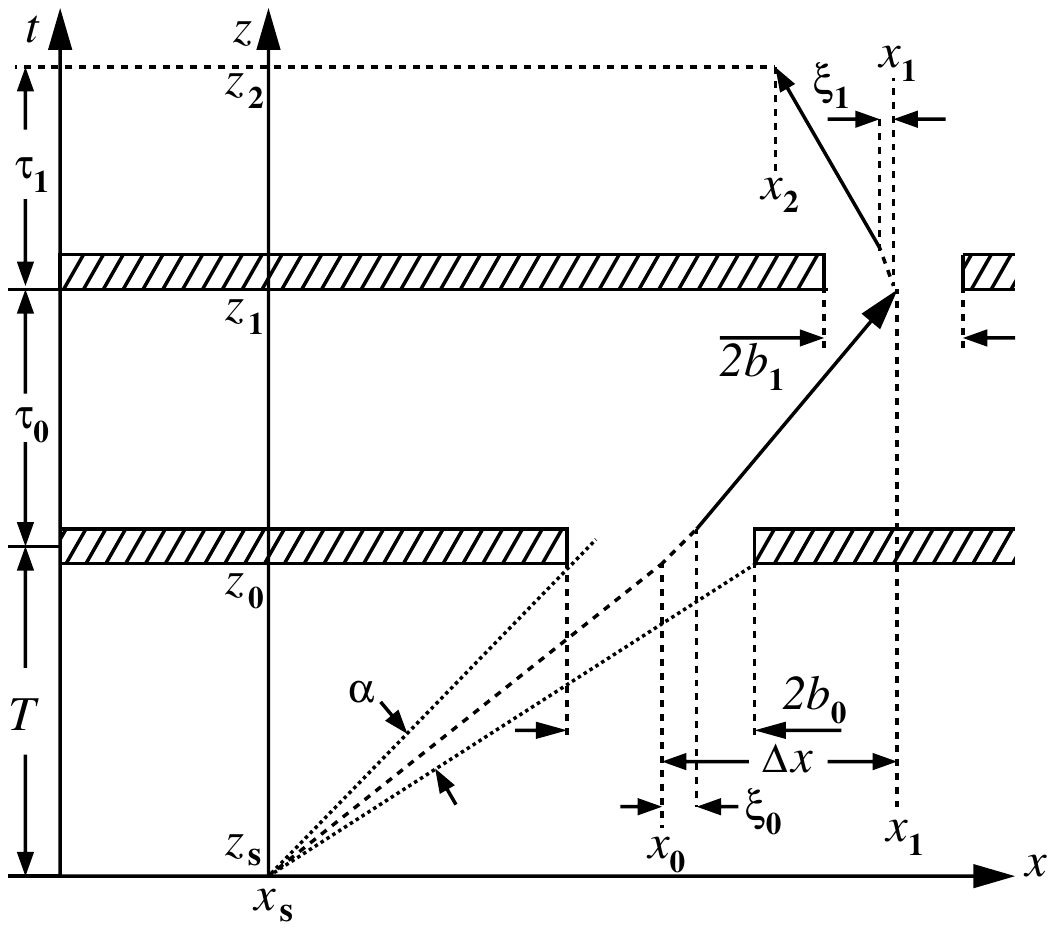}
  \end{picture}
  \caption{
  Passage of a particle along path
  $x_{\rm s}\rightarrow x_{0}\rightarrow x_{1}\rightarrow x_{2}$
  through two screens containing slits with widths
  $2b_{0}$ and $2b_{1}$.
  }
  \label{fig=3}
\end{figure}

 Computation of passing a particle through the system of two gratings, Fig.~\ref{fig=2}, is based on the path integral technique~\citep{FeynmanHibbs:1965, Ashmead1005}.
 Let us begin with writing the path integral, that describes passing the particle through slits made in two screens.
 They have been preliminarily prepared in two opaque screens
 situated perpendicularly to axis~$z$, see Fig.~\ref{fig=3}.
 For this reason we need to describe a movement of the particle between the screens
 and possible changing in its deflection at crossing the slits.
 We believe, that between the screens the particle moves as a free particle. Its Lagrangian is
\begin{equation}\label{eq=1}
    L = m\,{{\dot{x}^{\,2}}\over{2}} + {\rm const}.
\end{equation}
 Here $m$ is mass of the particle and $\dot{x}$ is its transversal velocity.
 A longitudinal momentum $p_{z}$ is much greater than its transverse component~\citep{ZeilengerEtAl2003, Hornberger2004, Hornberger2008}
 and we believe it is constant.

 The particles flying through the first slit along a ray $\alpha$ will pass this slit, ranging from $x_{0}-b_{0}$ to $x_{0}+b_{0}$, almost surely.
 And let its deflection on the first slit be such, that it passes through the second slit ranging from $x_{1}-b_{1}$ to $x_{1}+b_{1}$.
 The path integral in that case reads
\begin{eqnarray}
\nonumber 
 && \psi(x_{2},x_{1},x_{0},x_{\rm s}) \\
\nonumber
 &=& \int\limits_{-b_{1}}^{\;b_{1}}
  K(x_{2},T+\tau_{0}+\tau_{1};x_{1}+\xi_{1},T+\tau_{0}) \\
\nonumber
 && \int\limits_{-b_{0}}^{b_{0}} K(x_{1}+\xi_{1},T+\tau_{0};x_{0}+\xi_{0},T) \\
 &&\times K(x_{0}+\xi_{0},T;x_{\rm s},0)d\xi_{0}d\xi_{1}.
\label{eq=2}
\end{eqnarray}
 Integral kernels (propagators) for the particle freely flying are as follows~\citep{FeynmanHibbs:1965}
\begin{eqnarray}
\nonumber
  &&  K(x_{b},t_{b};x_{a},t_{a}) \\
  && \hspace{-24pt} = \Biggl[{{2\pi{\bf i}\hbar(t_{b}-t_{a})}\over{m}}\Biggr]^{-1/2}
  \exp\Biggl\{{{{\bf i}m(x_{b}-x_{a})^{2}}\over{2\hbar(t_{b}-t_{a})}}\Biggr\}.
\label{eq=3}
\end{eqnarray}
 By substituting the propagators into the path integral~(\ref{eq=2}) we get
\begin{widetext}
\begin{eqnarray}
\label{eq=4}
   && \psi(x_{2},x_{1},x_{0},x_{\rm s}) = \int\limits_{-b_{1}}^{\;b_{1}}
   \Biggl({{2\pi{\bf i}\hbar\tau_{1}}\over{m}}\Biggr)^{-1/2}
   \exp\Biggl\{{{{\bf i}m(x_{2}-(x_{1}+\xi_{1}))^{2}}\over{2\hbar\tau_{1}}}\Biggr\} \\
 \nonumber
 &&  \int\limits_{-b_{0}}^{b_{0}}
    \Biggl({{2\pi{\bf i}\hbar\tau_{0}}\over{m}}\Biggr)^{-1/2}
    \exp\Biggl\{{{{\bf i}m((x_{1}+\xi_{1})-(x_{0}+\xi_{0}))^{2}}\over{2\hbar\tau_{0}}}\Biggr\}
    \Biggl({{2\pi{\bf i}\hbar T}\over{m}}\Biggr)^{-1/2}
    \exp\Biggl\{{{{\bf i}m((x_{0}+\xi_{0})-x_{\rm s})^{2}}\over{2\hbar T}}\Biggr\}d\xi_{0}d\xi_{1}.
\end{eqnarray}
\end{widetext}
 The both integrals we can see are computed within finite intervals $[-b_{0},+b_{0}]$ and $[-b_{1},+b_{1}]$, respectively.
 Observe, that the integrating can be broadened from $-\infty$ to $+\infty$.
 But in this case we need to load the integrals by the step functions
 equal to unit within the finite intervals $[-b_{0},+b_{0}]$ and $[-b_{1},+b_{1}]$
 and they vanish outside of the intervals.

 In order to obtain exact solutions of the path integral let us take supposition, that the slits possess slightly fuzzy edges.
 In that case, the form factors, simulating the slits, can be represented by the Gaussian function~\citep{FeynmanHibbs:1965}
\begin{equation}\label{eq=5}
     G(\xi) = \exp\{-\xi^{2}/2b^{2}\}.
\end{equation}
 Effective width of the Gaussian curve is given by the parameter $b$.
 About two thirds of area of the curve are placed between points $-b$ and $+b$,
 and one third of the area is beyond the interval $(-b,+b)$.
 Consequently, tunneling through the screen in the vicinity of the slits we believe can take place.
 It is due to the fuzzy edges adopted above.

 Let us insert into the integrals the Gaussian factors $G(\xi_{0})$ and $G(\xi_{1})$ with the parameters $b_{0}$ and $b_{1}$
 and replace the finite limits $\pm b_{0}$ and $\pm b_{1}$ by the limits from $-\infty$ to $+\infty$.
 Take into account, that we have a solution for the inner integral, that has been written out in~\citep{Sbitnev1001}.
 Now we can rewrite the integral~(\ref{eq=4}) more definitely
\begin{widetext}
\begin{eqnarray}\nonumber
 && 
 \psi(x_{2},x_{1},x_{0},x_{\rm s}) = 
  \Biggl({{1}\over{T}}+{{1}\over{\tau_{0}}}+{{{\bf i}\hbar}\over{mb_{0}^{2}}}\Biggr)^{-1/2}
  \int\limits_{-\infty}^{\infty}
  {{m G(\xi_{1})}\over{2\pi{\bf i}\hbar\sqrt{T\tau_{0}\tau_{1}}}}
  \exp\Biggl\{{{{\bf i}m(x_{2}-(x_{1}+\xi_{1}))^{2}}\over{2\hbar\tau_{1}}}\Biggr\} \\
  && \hspace{-18pt}
  \times \exp\Biggl\{{{{\bf i}m}\over{2\hbar}} \Biggl(
  \Biggl({{((x_{1}+\xi_{1})-x_{0})^{2}}\over{\tau_{0}}}+{{(x_{0}-x_{\rm s})^{2}}\over{T}}
  \Biggr) -
  {{(-((x_{1}+\xi_{1})-x_{0})/\tau_{0}+(x_{0}-x_{\rm s})/T)^{2}}\over
  {(1/\tau_{0}+1/T+{\bf i}\hbar/mb_{0}^{2})}} \Biggr)
   \Biggr\}d\xi_{1}.
\label{eq=6}
\end{eqnarray}
\end{widetext}
 Solutions of such integrals stem from the formula
\begin{equation}\label{eq=7}
    \int\limits_{-\infty}^{\infty}
    {\rm e}^{\alpha\,\xi^{2}+\beta\,\xi+\gamma}\,d\xi
    = \sqrt{{{\pi}\over{-\alpha}}}\,{\rm e}^{-\beta^{2}/4\alpha+\gamma}.
\end{equation}
 In order to get such a form,
 let us regroup all terms under the integral~(\ref{eq=6}) and collect their
 at coefficients $\xi_{1}^{2}$, $\xi_{1}$, and free from it.
 We have
\begin{widetext}
 \begin{enumerate}
   \item the term at $\xi_{1}^{2}$:
\begin{equation}\label{eq=8}
   \alpha = {{{\bf i}m}\over{2\hbar}}
    \Biggl(
    {{1}\over{\tau_{1}}} + {{1}\over{\tau_{0}}}
    + {{{\bf i}\hbar}\over{m b_{1}^{2}}}
    -{{1}\over{\tau_{0}^{2}(1/\tau_{0}+1/T+{\bf i}\hbar/mb_{0}^{2})}}
    \Biggr);
\end{equation}
   \item the term at $\xi_{1}$:
\begin{equation}\label{eq=9}
    \beta = -2\,{{{\bf i}m}\over{2\hbar}} \Biggl(
     {{(x_{2}-x_{1})}\over{\tau_{1}}} - {{(x_{1}-x_{0})}\over{\tau_{0}}}
     + {{(x_{1}-x_{0})/\tau_{0}^{2}-(x_{0}-x_{\rm s})/\tau_{0}T}\over
     {(1/\tau_{0}+1/T+{\bf i}\hbar/mb_{0}^{2})}}
    \Biggr);
\end{equation}
   \item the term free from $\xi_{1}$:
\begin{equation}\label{eq=10}
    \gamma = {{{\bf i}m}\over{2\hbar}} \Biggl(
    {{(x_{2}-x_{1})^{2}}\over{\tau_{1}}} + {{(x_{1}-x_{0})^{2}}\over{\tau_{0}}} + {{(x_{0}-x_{\rm s})^{2}}\over{T}}
    - {{((x_{1}-x_{0})/\tau_{0}-(x_{0}-x_{\rm s})/T)^{2}}\over{(1/\tau_{0}+1/T+{\bf i}\hbar/mb_{0}^{2})}}
    \Biggr).
\end{equation}
 \end{enumerate}
\end{widetext}

 Now we can express the terms $(\pi/(-\alpha))^{1/2}$ and $\gamma-\beta/4\alpha$ in the right part of Eq.~(\ref{eq=7}).
 The first term, accurate to the multiplicand
\begin{equation}\label{eq=11}
    \Biggl({{1}\over{T}}+{{1}\over{\tau_{0}}}+{{{\bf i}\hbar}\over{mb_{0}^{2}}}\Biggr)^{-1/2}
  {{m}\over{2\pi{\bf i}\hbar\sqrt{T\tau_{0}\tau_{1}}}},
\end{equation}
 relates to an amplitude factor $A$ of the wave function $\psi(x_{2},x_{1},x_{0},x_{\rm s})$.
 For this reason, by multiplying the term $(\pi/(-\alpha))^{1/2}$ by the multiplicand~(\ref{eq=11})
 we obtain the amplitude factor $A$. It has the following view
\begin{widetext}
\begin{equation}
    A =
 \sqrt{{{m}\over{2\pi{\bf i}\hbar\,T}}} \cdot
 {{1}\over{
  \sqrt{
 \displaystyle
  \Biggl(1+{{\tau_{1}}\over{\tau_{0}}}\Biggr)
  \Biggl({\displaystyle
  1 + {{{\bf i}\hbar\tau_{1}}\over{mb_{1}^{2}(1+\tau_{1}/\tau_{0})}}
  }\Biggr)
  \Biggl(1+{{\tau_{0}}\over{T}}\Biggr)
  \Biggl({\displaystyle
  1 + {{{\bf i}\hbar\tau_{0}}\over{mb_{0}^{2}(1+\tau_{0}/T)}}
  }\Biggr)
  - {{\tau_{1}}\over{\tau_{0}}} }
         }}.
\label{eq=12}
\end{equation}
 In turn, the term $\gamma-\beta^{2}/4\alpha$ reads:
\begin{eqnarray}
\nonumber 
  \gamma-\beta^{2}/4\alpha =
  {{{\bf i}m}\over{2\hbar}} \hspace{-8pt}
&&
  \Biggl[
  \Biggl(
    {{(x_{2}-x_{1})^{2}}\over{\tau_{1}}} + {{(x_{1}-x_{0})^{2}}\over{\tau_{0}}} + {{(x_{0}-x_{\rm s})^{2}}\over{T}}
    - {{((x_{1}-x_{0})/\tau_{0}-(x_{0}-x_{\rm s})/T)^{2}}\over{((\tau_{0}+T)/T\tau_{0}+{\bf i}\hbar/mb_{0}^{2})}}
    \Biggr) \\
 &&
   -{{\displaystyle\Biggl({{(x_{2}-x_{1})}\over{\tau_{1}}} - {{(x_{1}-x_{0})}\over{\tau_{0}}}
     + {{(x_{1}-x_{0})/\tau_{0}-(x_{0}-x_{\rm s})/T}\over
     {\tau_{0}((\tau_{0}+T)/T\tau_{0}+{\bf i}\hbar/mb_{0}^{2})}}
    \Biggr)^{2}}
  \over{\displaystyle
 \Biggl(
  {\displaystyle{{\tau_{0}+T}\over{T\tau_{0}}} + {{{\bf i}\hbar}\over{mb_{0}^{2}}}}
 \Biggr)^{-1}
 \Biggl(
 \Biggl(\displaystyle
 {{\tau_{1}+\tau_{0}}\over{\tau_{1}\tau_{0}}} + {{{\bf i}\hbar}\over{mb_{1}^{2}}} \Biggr)
 \Biggl(
  {\displaystyle{{\tau_{0}+T}\over{T\tau_{0}}} + {{{\bf i}\hbar}\over{mb_{0}^{2}}}}
 \Biggr)
  - {{1}\over{\tau_{0}^{2}}}
 \Biggr)
 }}
    \hspace{-6pt}
    \left. \matrix{\cr\cr\cr\cr\cr}\right]
\label{eq=13}
\end{eqnarray}
\end{widetext}

\subsection{\label{subsec:level2A}Series of replacements}

 Let us now define effective slit's half-widths
\begin{equation}\label{eq=14}
    \sigma_{j,0} = {{b_{j}}\over{\sqrt{2}}},
\end{equation}
 where $j$ takes numbers $0$ and $1$.
 And next we determine a complex time-dependent spreading
\begin{eqnarray}\label{eq=15}
    \sigma_{j,\tau_{j}}&=&\sigma_{j,0} + {\bf i}\, {{\hbar\tau_{j}}\over{2m \sigma_{j,0}(1+\tau_{j}/\tau_{j-1})}}
\end{eqnarray}
 where $j=0,1$ and at $j=0$ we suppose $\tau_{-1}=T$.
 It should be noted, the complex spreading parameters have been presented in works~\citep{SanzMiret2007, SanzMiret2008}.
 Here we choose the same representation.

 More one step is to replace flight times $T$, $\tau_{0}$, and $\tau_{1}$ by flight distances $(z_{0}-z_{\rm s})$, $(z_{1}-z_{0})$, and $(z_{2}-z_{1})$, Fig.~\ref{fig=3}.
 This replacement reads
\begin{equation}\label{eq=16}
    \left\{
      \begin{array}{c}
        T        = (z_{0}-z_{\rm s})/v_{z}, \\
        \tau_{0} = (z_{1}-z_{0})/v_{z}, \\
        \tau_{1} = (z_{2}-z_{1})/v_{z}, \\
      \end{array}
    \right.
\end{equation}
 where $v_{z}$ is a particle velocity along the axis~$z$.
 We note that $mv_{z}=p_{z}$ is $z$-component of the particle momentum.
 Next, we introduce the de Broglie wavelength  $\lambda_{_{\rm dB}}=h/p_{z}$,
 where $h=2\pi\hbar$ is the Planck constant.
 Rewrite the complex spreading~(\ref{eq=15}) according to these remarks
\begin{eqnarray}
     \sigma_{j,\tau_{j}\rightarrow z_{j}}
     &=& \sigma_{j,0} + {\bf i}\,
     {{\lambda_{_{\rm dB}}(z_{j+1}-z_{j})}\over
     {4\pi\sigma_{j,0}{\displaystyle\Biggl({{z_{j+1}-z_{j-1}}\over{z_{j}-z_{j-1}}}\Biggr)}}}.\hspace{12pt}
\label{eq=17}
\end{eqnarray}
 Hereinafter, for brevity, we will not write the subscript~dB at $\lambda$.

 Define a dimensionless complex distance-dependent spreading as follows
\begin{eqnarray}
\nonumber
  \Sigma_{j,z_{j}}  &=& \Biggl({{z_{j+1}-z_{j-1}}\over{z_{j}-z_{j-1}}}\Biggr) {{\sigma_{j,z_{j}}}\over{\sigma_{j,0}}} \\
   &=& {{z_{j+1}-z_{j-1}}\over{z_{j}-z_{j-1}}}
   +
   {\bf i}
    {{\lambda(z_{j+1}-z_{j})}\over{4\pi\sigma_{j,0}^{\,2}}}.
\label{eq=18}
\end{eqnarray}
 Also we define a dimensionless parameter
\begin{equation}\label{eq=19}
      \Xi_{\,0} = 1 -{{(x_{0}-x_{\rm s})}\over{(z_{0}-z_{\rm s})}}{{(z_{1}-z_{0})}\over{(x_{1}-x_{0})}}
\end{equation}
 that tends to 1 as $z_{\rm s}\rightarrow-\infty$.
 Now we can rewrite the terms $A$ and $\gamma-\beta^{\,2}/4\alpha$, represented in Eqs.~(\ref{eq=12}) and~(\ref{eq=13}),
 by replacing cumbersome parameters by defined above.

 The amplitude factor, $A$, of the wave function $\psi(x_{2},x_{1},x_{0},x_{\rm s})$ has the following view
\begin{equation}\label{eq=20}
    A =
     \sqrt{{{m}\over{2\pi{\bf i}\,\hbar\,T}}} \cdot
 {{1}\over{{
 D(\Sigma_{0,z_{0}},\Sigma_{1,z_{1}})
 }}}.
\end{equation}
 Here we do not replace $T$ by $(z_{0}-z_{\rm s})/v_{z}$.
 It points out to distance to the source accurate to division by  $v_{z}$.
 As we remove the sources to infinity ($T\rightarrow\infty$) the amplitude factor tends to zero.
 Nevertheless, we keep a finite value of the parameter $A$ as a normalization factor of the wave function~\citep{Sbitnev1001}.

 The phase term, $\gamma-\beta^{2}/4\alpha$
 for the same wave function $\psi(x_{2},x_{1},x_{0},x_{\rm s})$ reads
\begin{widetext}
\begin{eqnarray}
\nonumber
  \gamma-\beta^{2}/4\alpha &=& {{\bf i}\pi}\,
    \Biggl[
    \Biggl(
    {{(x_{2}-x_{1})^{2}}\over{\lambda(z_{2}-z_{1})}}
    + {{(x_{1}-x_{0})^{2}}\over{\lambda(z_{1}-z_{0})}}
   \Biggl(
    1 -
    {{\Xi_{\,0}^{\,2}}\over{\Sigma_{0,z_{0}}}}
    \Biggr)
    + {{(x_{0}-x_{\rm s})^{2}}\over{\lambda(z_{0}-z_{\rm s})}}
    \Biggr)
    \\
  && \hspace{-14pt}
-
{{\displaystyle
 {{\lambda(z_{2}-z_{1})
 \Sigma_{0,z_{0}}
 }}
 \over { D(\Sigma_{0,z_{0}},\Sigma_{1,z_{1}})^{2}
}}
}
\Biggl(
     {{(x_{2}-x_{1})}\over{\lambda(z_{2}-z_{1})}} -
     {{(x_{1}-x_{0})}\over{\lambda(z_{1}-z_{0})}}
\Biggl(
     1 -
 {{\Xi_{\,0}}\over{\Sigma_{0,z_{0}}}}
\Biggr)
\Biggr)^{2}\,
\Biggr],
\label{eq=21}
\end{eqnarray}
\end{widetext}
 The term $D(\Sigma_{0,z_{0}},\Sigma_{1,z_{1}})$ represented in divisors of expressions~(\ref{eq=20}) and~(\ref{eq=21})
 is as follows
\begin{equation}\label{eq=22}
    D(\Sigma_{0,z_{0}},\Sigma_{1,z_{1}}) =
    \sqrt{
    \Sigma_{0,z_{0}}\Sigma_{1,z_{1}} -
    {{z_{2}-z_{1}}\over{z_{1}-z_{0}}}
    }.
\end{equation}

\subsection{\label{subsec:level2B}Matter waves behind the gratings $G_{0}$ and $G_{1}$}

 For observation of the wave field behind the second grating, $G_{1}$, we need to situate a detector in a point $(x_{2},z_{2})$, Fig.~\ref{fig=3}.
 On the other hand, if we wish to observe the wave field between the first and second gratings, between $G_{0}$ and $G_{1}$,
 we must situate the detector in a point $(x_{1},z_{1})$.
 In order to realize the second case, it is sufficient to put in the expressions~(\ref{eq=21}) and~(\ref{eq=22}) $x_{2}=x_{1}$ and $z_{2}=z_{1}$.
 Let now the point $(x_{2},z_{2})=(x_{1},z_{1})$ be situated in a region between the gratings.
 Observe, in this case, that all terms containing differences $(z_{2}-z_{1})$ vanish.
 Next, we move the detector in the point $(x_{1},z_{1})\rightarrow(x,z)$ that can be situated anywhere between the gratings.

 Thus, we will have in mind that the expressions~(\ref{eq=21}) and~(\ref{eq=22}) contain full information about wave fields both between the gratings and behind the second grating. It depends on choosing coordinates of position of the detector, either $z_{2}\rightarrow z$, $x_{2}\rightarrow x$, or $z_{2}=z_{1}\rightarrow z$, $x_{2}=x_{1}\rightarrow x$.
 We write variables $(x,z)$ instead of $(x_{2},z_{2})$ at describing the wave field behind the second grating.
 And we write the same variables $(x,z)$ instead of  $(x_{1},z_{1})$ when we describe the wave pattern between the gratings.
 In the last case, we have replacements $(z_{2}-z_{1})\rightarrow(z-z_{1})\rightarrow(z-z)=0$, $(x_{2}-x_{1})\rightarrow(x-x_{1})\rightarrow(x-x)=0$.
 Terms containing such differences either vanish, or become unit if a ratio $(x-x)/(z-z)$ takes place.
 Now we can write out the wave patterns at passing the particle through single slit within each grating.

 A particle emitted from the point $(x_{\rm s},z_{\rm s})$, belonging to the source, can be detected
 in the point $(x_{2},z_{2})\Rightarrow(x_{},z_{})$, $z>z_{1}$ (behind the second grating) in accordance with
 the following wave function
\begin{widetext}
\begin{eqnarray}
\nonumber
  \psi(x,z,x_{1},x_{0},x_{\rm s}) &=& {\displaystyle{{\sqrt{\displaystyle{{m}\over{2\pi{\bf i}\hbar T}}}}\over{D(\Sigma_{0,z_{0}},\Sigma_{1,z_{1}})}}}
   \exp\Biggl\{
   {{\bf i}\pi}\,
    \Biggl[
    \Biggl(
    {{(x_{}-x_{1})^{2}}\over{\lambda(z_{}-z_{1})}}
    + {{(x_{1}-x_{0})^{2}}\over{\lambda(z_{1}-z_{0})}}
   \Biggl(
    1 -
    {{\Xi_{\,0}^{\,2}}\over{\Sigma_{0,z_{0}}}}
    \Biggr)
    + {{(x_{0}-x_{\rm s})^{2}}\over{\lambda(z_{0}-z_{\rm s})}}
    \Biggr)
    \\
  && \hspace{-14pt}
-
{{\displaystyle
 {{\lambda(z_{}-z_{1})
 \Sigma_{0,z_{0}}
 }}
 \over { D(\Sigma_{0,z_{0}},\Sigma_{1,z_{1}})^{2}
}}
}
\Biggl(
     {{(x_{}-x_{1})}\over{\lambda(z_{}-z_{1})}} -
     {{(x_{1}-x_{0})}\over{\lambda(z_{1}-z_{0})}}
\Biggl(
     1 -
 {{\Xi_{\,0}}\over{\Sigma_{0,z_{0}}}}
\Biggr)
\Biggr)^{2}\,
\Biggr] \Biggr\}
\label{eq=23}
\end{eqnarray}
 In the region between the gratings, $z<z_{1}$, we employ replacement $(x_{1},z_{1})\Rightarrow(x_{},z_{})$ and consequently
\begin{equation}\label{eq=24}
    \psi(x,z,x_{0},x_{\rm s}) = {\displaystyle{\sqrt{\displaystyle{{m}\over{2\pi{\bf i}\hbar T\Sigma_{0,z_{0}}}}}}}\cdot
   \exp\Biggl\{
   {{\bf i}\pi}\,
    \Biggl[
     {{(x_{}-x_{0})^{2}}\over{\lambda(z_{}-z_{0})}}
   \Biggl(
    1 -
    {{\Xi_{\,0}^{\,2}}\over{\Sigma_{0,z_{0}}}}
    \Biggr)
    + {{(x_{0}-x_{\rm s})^{2}}\over{\lambda(z_{0}-z_{\rm s})}}
    \Biggr] \Biggr\}
\end{equation}
\end{widetext}
 As we stated above, the function $\psi(x,z,x_{0},x_{\rm s})$ stems from $\psi(x,z,x_{1},x_{0},x_{\rm s})$
 as soon as we put in Eq.~(\ref{eq=23}) values $z_{1}=z$ and $x_{1}=x$.
 Observe, that $D(\Sigma_{0,z_{0}},\Sigma_{1,z_{1}})=\Sigma_{0,z_{0}}^{1/2}$ so far as $\Sigma_{1,z_{1}}=1$ at $z=z_{1}=z_{2}$, see Eqs.~(\ref{eq=18}) and~(\ref{eq=22}).

\section{\label{sec:level3}Interference patterns from two-grating structure}

 The gratings shown in Fig.~\ref{fig=2} we believe have different amount of slits.
 The first grating has $N_{0}$ slits, so $n_{0}=0,1,\cdots,(N_{0}-1)$.
 The second grating has $N_{1}$ slits, and $n_{1}=0,1,\cdots,(N_{1}-1)$.
 Distance between slits for the first grating is $d_{0}$ and for the second grating is $d_{1}$.
 Now we can write out a wave field beyond the gratings
 that is composed of superposition of wave functions
 representing coherent emission from all slits.

 The matter wave in a zone between the first and second  gratings reads
\begin{eqnarray}
\nonumber
   &&
   |\Psi_{0}(x,z,x_{\rm s},\lambda_{})\rangle = \\
   && 
    \sum_{n_{0}=0}^{N_{0}-1}
    \psi\Biggl(x,z,\biggl({ n_{0}-{{N_{0}-1}\over{2}}}\biggr)d_{0},x_{\rm s}\Biggr).
\label{eq=25}
\end{eqnarray}
 And the matter wave extending beyond the second grating reads
\begin{eqnarray}
 \nonumber 
  && |\Psi_{1}(x,z,x_{\rm s},\lambda_{})\rangle = \\ &&
  \sum_{n_{1}=0}^{N_{1}-1}\sum_{n_{0}=0}^{N_{0}-1}
  \psi\Biggl(x,z,
  \biggl({\small n_{1}-{{N_{1}-1}\over{2}}}\biggr)d_{1},   \nonumber \\ && \hspace{64pt}
  \biggl({\small n_{0}-{{N_{0}-1}\over{2}}}\biggr)d_{0},x_{\rm s}
  \Biggr).
\label{eq=26}
\end{eqnarray}
 The both wave functions are represented without normalized factors.
 We need no here in these factors, since our interest is to show a general pattern of the density distribution
\begin{equation}\label{eq=27}
 \hspace{0pt}
    p(x,z)=\langle \Psi(x,z,x_{\rm s},\lambda_{})|\Psi(x,z,x_{\rm s},\lambda_{}) \rangle.
\end{equation}

 As for the particles supporting the matter wave here, we consider fullerene molecules~\citep{JuffmannEtAl2010}.
 They are massive molecules. Its mass is about $m_{\rm C_{60}}\approx1.2\times10^{-24}$~kg.
 The fullerene molecule has radius $R_{\rm C_{60}}=350$~pm \citep{YanovLeszczynski2004}.
 In turn, the de Broglie wavelength is much smaller than this radius.
 For instance, the de Broglie wavelength equal to $5$ pm appears in the experimental work~\citep{JuffmannEtAl2010}.
 The wavelength $\lambda_{\rm dB}=5$~pm is adopted in this work as well.
 The widths of the slits (open slit windows) are as small as $2b_{0}= 75$~nm in $G_{0}$ and $2b_{1}=150$~nm in $G_{1}$.
 Distances between the slits are equivalent for the both gratings $d_{0}=d_{1}=500$~nm.
 The distances have been double increased in contrast to those given in~\citep{JuffmannEtAl2010} and~\citep{Sbitnev1001}.
 It is done in order that we could see definitely the Talbot carpets arising between the gratings, and maybe beyond the both gratings.

\begin{figure}[htb!]
  \centering
  \begin{picture}(200,540)(25,15)
      \includegraphics[scale=0.8]{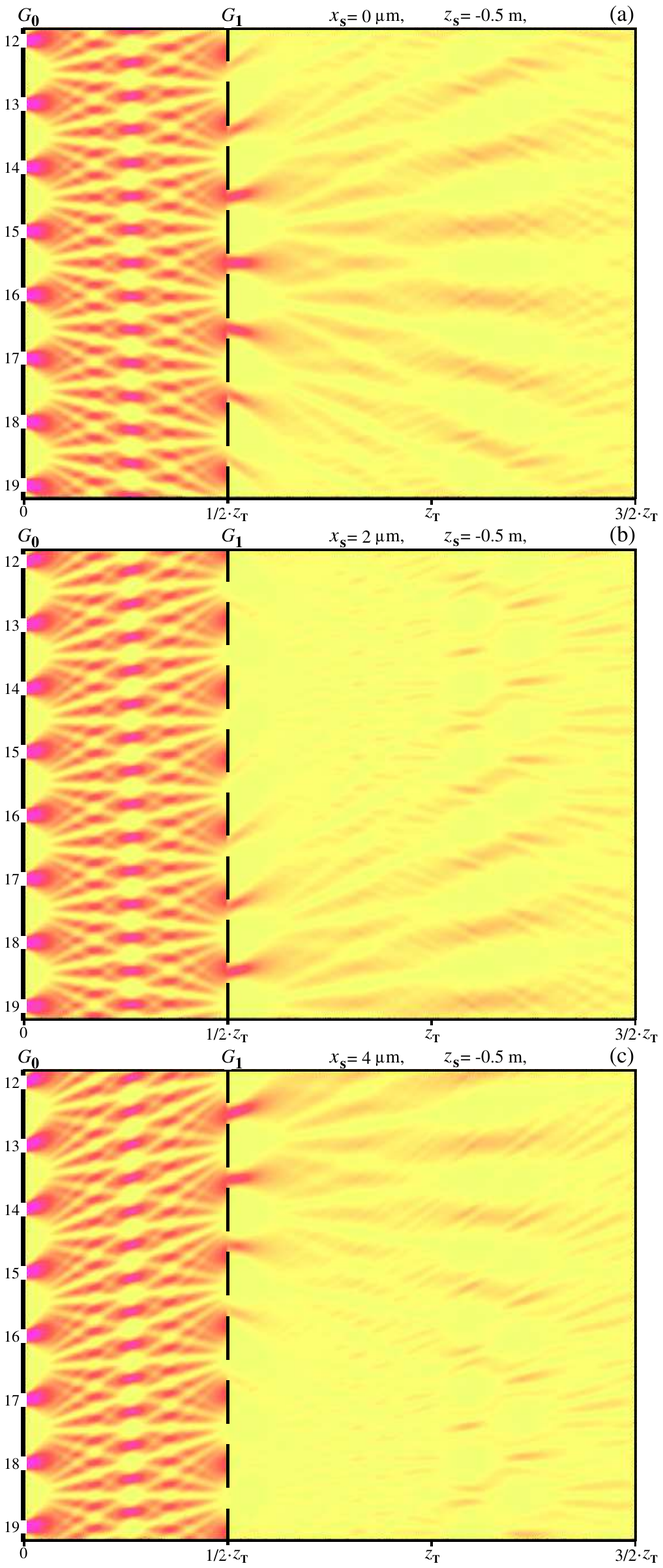}
  \end{picture}
  \caption{
  Density distribution pattern $p(x,z)$
  in the near-field region $z\in(0,1.5 z_{\rm T})=(0,0.15)$ m, $N_{0}=32$, $N_{1}=33$,
  de Broglie wavelength $\lambda_{\rm dB}=5$ pm, Talbot length $z_{\rm T}=0.1$~m.
  Distance from $G_{0}$ to the source is $z_{\rm s}=-0.5$ m :
  (a)~the source is situated on the optical axis, $x_{\rm s}=0$;
  (b)~the source is shifted from the central axis on $x_{\rm s}=2~\mu$m;
  (c)~the source is shifted from the central axis on $x_{\rm s}=4~\mu$m.
  }
  \label{fig=4}
\end{figure}

 Here we consider example of emergence of interference patterns from the gratings $G_{0}$ and $G_{1}$ containing even amount of slits, $N_{0}=32$,
 and odd amount of slits, $N_{1}=33$. Difference in parity of the numbers $N_{0}$ and $N_{1}$ is conditioned by the fact, that the first self-image
 of $G_{0}$ is shifted  exactly on half period of the grating. Due to this trick, the slits of $G_{1}$ are located exactly on nodes of the first self-image of the grating $G_{0}$.
 The grating $G_{1}$ keeps "open gates" for particles to spread further.

 Fig.~\ref{fig=4} shows the density distribution $p(x,z)$ in the near-field region
 for different positions of a point source:
 (a)~$x_{\rm s} = 0~\mu$m; (b)~$x_{\rm s} = 2~\mu$m; and (c)~$x_{\rm s} = 4~\mu$m,
 and at $z_{\rm s}$ distant from the first grating on $-0.5$ m.
 Here we have shown cases of shifting the point source to a positive area, $x_{s}>~0$.
 As for negative shifting it can be obtained by simple reflection of the interference patterns
 about the plane $(y, z)$ intersecting axis $x$ at $x=~0$, see frame of axis in Fig.~\ref{fig=2}.
 That is, the slits should be subjected to the following inversion
 $12\leftrightarrow19$, $13\leftrightarrow18$, $14\leftrightarrow17$, $15\leftrightarrow16$.

 We can see in Figs.~\ref{fig=4}(a),~~\ref{fig=4}(b), and~\ref{fig=4}(c) ripples of high order against the background of the basic divergent rays.
 They are induced  by presence of many lateral slits invisible in these figures.
 The interference  patterns are seen to change at changing position of the point source.
\begin{figure}[htb!]
  \centering
  \begin{picture}(200,540)(25,15)
      \includegraphics[scale=0.8]{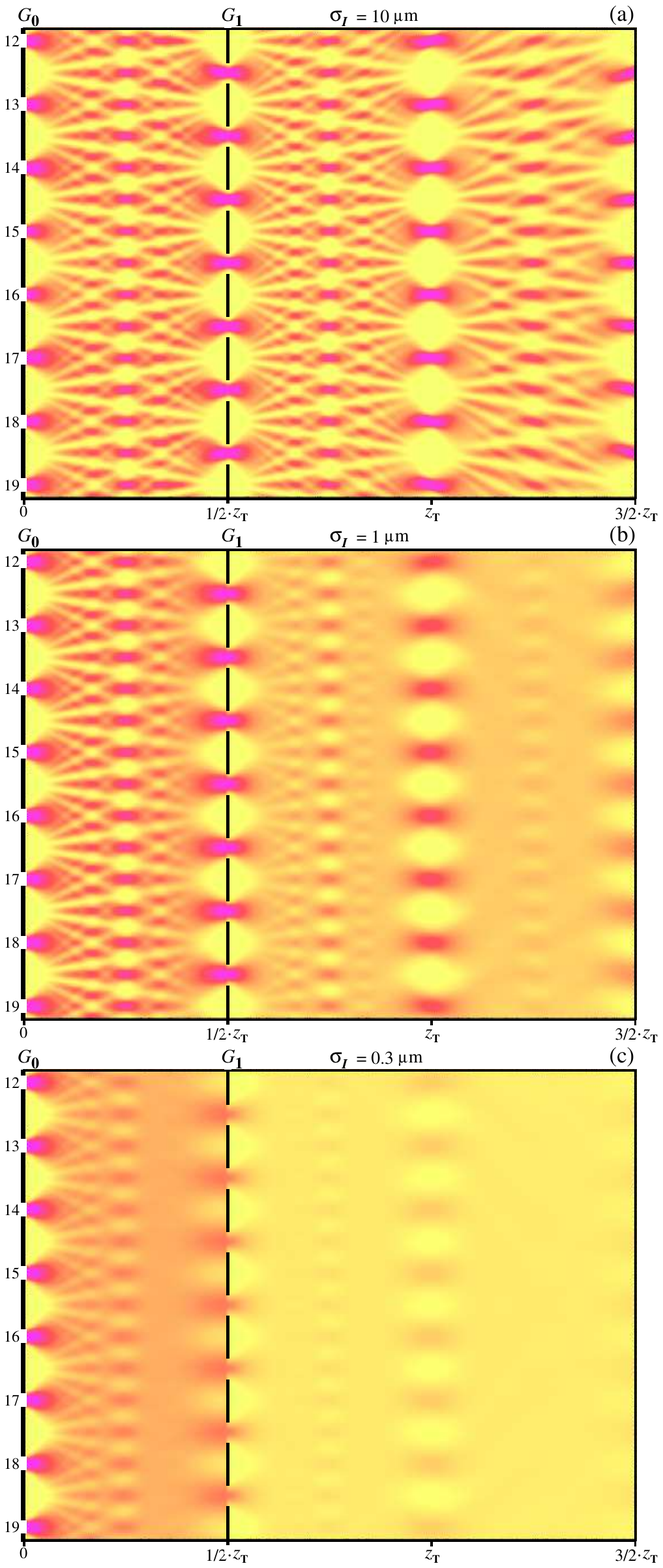}
  \end{picture}
  \caption{
 Density distribution pattern $p(x,z)$ averaged over all point sources localized with increment ${\delta x_{\rm s}}=0.25~\mu$m  in the interval $x_{\rm s}\in[-4, 4]~\mu$m, $z_{\rm s}=-0.5$ m, $N_{0}=32$, $N_{1}=33$.
 The averaging has been carried out with the Gaussian kernel~(\ref{eq=29}) loaded by the dispersion constant $\sigma_{I}$:
 (a)~coherent beam,~$\sigma_{_I}=10~\mu$m; (b)~almost coherent beam,~$\sigma_{_I}=1~\mu$m; (c)~almost noncoherent beam,~$\sigma_{_I}=0.3~\mu$m.
  }
  \label{fig=5}
\end{figure}

\subsection{\label{subsec:level3A}Coherent and noncoherent sources}

 Let us now turn to the coherence properties of the particle beam~\citep{ZeilengerEtAl2003}.
 We will assume, that point sources are coherent,
 if wave functions relating to these sources are summed together in the vicinity of a detector.
 And they are noncoherent
 if intensities (probability densities) are summed.
 In particular, superposition of the wave functions in the vicinity of the detector can be with weakened contribution of cross terms,
 that provide interference effects.
 This case relates to intermediate variants, which can be considered by appealing to
 the Gaussian Schell-model~\citep{MandelWolf1995, GburWolf2001}.
 We have to keep in mind, however, that the wave functions are primary quantum subjects,
 whereas intensities are found on a final stage at reading from detectors.

 Averaging of the wave functions from all point sources distributed along $x_{\rm s}$ at fixed $z_{\rm s}$ is carried out in the following form:
\begin{eqnarray}
\nonumber
&&
\hspace{28pt}
    p(x,z) = \sigma_{_I}\sum\limits_{x_{\rm s}^{\,'}}\sum\limits_{x_{\rm s}^{\,''}} \\
&&
\hspace{-28pt}
    \langle \Psi(x,z,x_{\rm s}^{\,'},\lambda_{})|
   \mu(x_{\rm s}^{\,'},x_{\rm s}^{\,''},\sigma_{_I})|\Psi(x,z,x_{\rm s}^{\,''},\lambda_{}) \rangle.
\label{eq=28}
\end{eqnarray}
 The Gaussian kernel $\mu(x_{\rm s}^{\,'},x_{\rm s}^{\,''},\sigma_{I})$ reads
\begin{equation}\label{eq=29}
\hspace{-8pt}
    \mu(x_{\rm s}^{\,'},x_{\rm s}^{\,''},\sigma_{_I})
    = {{1}\over{\sqrt{2\pi}\,\sigma_{_I}}}
    \exp\biggl\{
    -{{(x_{\rm s}^{\,'}-x_{\rm s}^{\,''})^{2}}\over{2\,\sigma^{\,2}_{_I}}}
    \biggr\}
\end{equation}
 with the dispersion parameter $\sigma_{_I}$ being an effective coherent width of the beam.
 This parameter presented as a factor in front of the sums~(\ref{eq=28}) provides identity of dimensionalities
 for the probability density distributions.

 First, one can notice that at $\sigma_{_I}\ll 1$, the Gaussian kernel fits the Dirac $\delta$-function.
 And the expression~(\ref{eq=28}) drops to a simple summation of the probability densities
\begin{equation}\label{eq=30}
    p(x,z) \sim \sum\limits_{x_{\rm s}^{}}
     \langle \Psi(x,z,x_{\rm s}^{},\lambda_{})|\Psi(x,z,x_{\rm s}^{},\lambda_{}) \rangle.
\end{equation}
 And at $\sigma_{_I}\gg 1$ the Gaussian kernel degenerates to a constant. In that case we have
\begin{eqnarray}
\nonumber
   && p(x,z) \sim \sum\limits_{x_{\rm s}^{}}
     \langle \Psi(x,z,x_{\rm s}^{},\lambda_{})|\Psi(x,z,x_{\rm s}^{},\lambda_{}) \rangle \\
   &&\hspace{-18pt}
    + \sum\limits_{x_{\rm s}^{\,'}\ne x_{\rm s}^{\,''}}\sum\limits_{x_{\rm s}^{\,''}}
   \langle \Psi(x,z,x_{\rm s}^{\,'},\lambda_{})|\Psi(x,z,x_{\rm s}^{\,''},\lambda_{}) \rangle.
\label{eq=31}
\end{eqnarray}
 The second sum here contains the cross terms, that introduce interference effects from different point sources.

 The expression~(\ref{eq=30}) represents an example of wholly noncoherent beam.
 Whereas, the expression~(\ref{eq=31}) gives completely coherent beam.
 Intermediate coherence beams are possible as well.

 Figs.~\ref{fig=5}(a),~\ref{fig=5}(b), and~\ref{fig=5}(c) show averaged density distributions for different depth of coherence
 ranging from a coherent beam to noncoherent one.
 Summation is taken for all point sources localized at $x_{\rm s}=-4, -3.25, -3.5,\cdots,4~\mu$m, the increment is $\delta x_{\rm s}=0.25~\mu$m.
 Distance to the source is $z_{\rm s}=-0.5$~m.
 The figures demonstrate interference patterns that are reproduced from (a) coherent beam, $\sigma_{_I}=10~\mu$m,
 to (c) almost noncoherent beam, $\sigma_{_I}=0.3~\mu$m.
 The high-order interference fringes are seen to be washed out
 as the dispersion parameter $\sigma_{_I}$ decreases from $10~\mu$m to $0.1~\mu$m,
 what is in good agreement with computational results given in~\citep{CroninMcMorran2006}.
\begin{figure}[htb!]
  \centering
  \begin{picture}(200,230)(20,10)
      \includegraphics[scale=0.8]{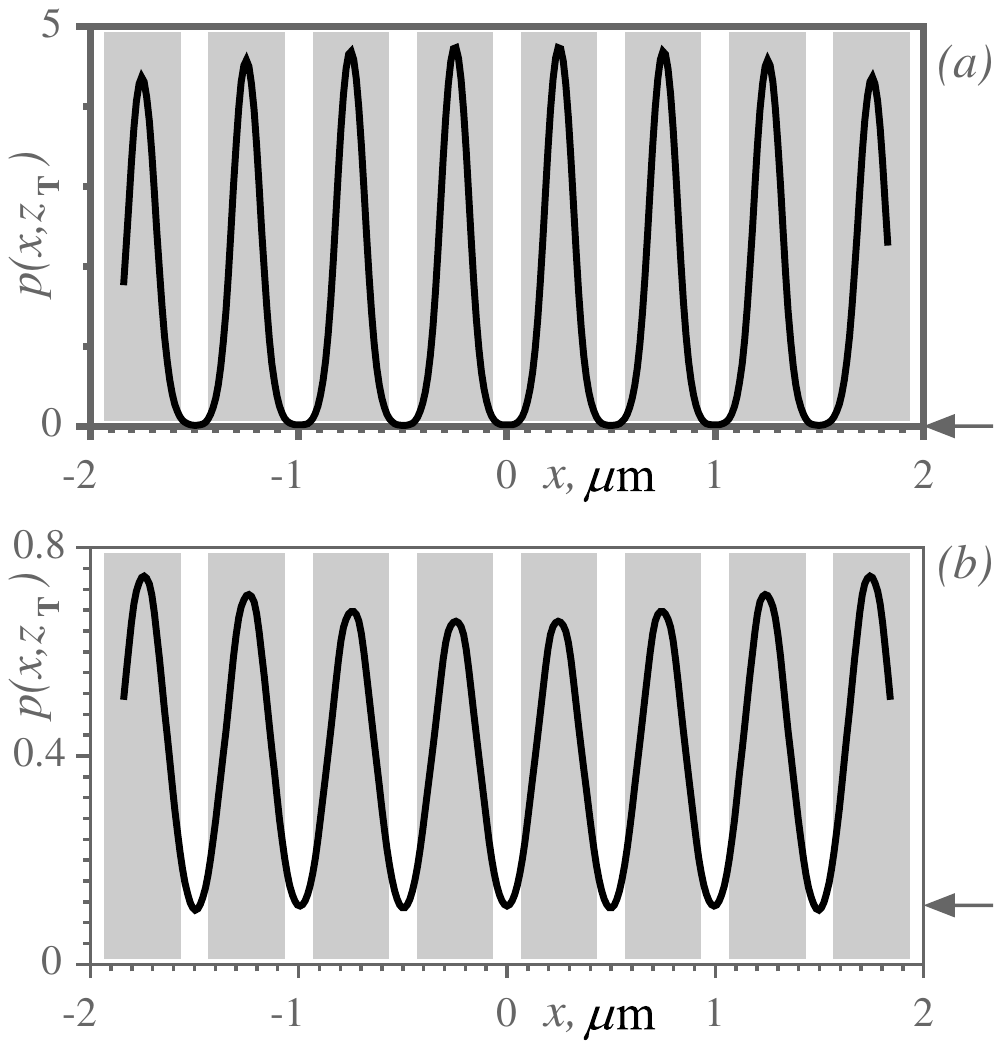}
  \end{picture}
  \caption{
  Interference fringes observed in the cross-section of the density distribution pattern~(\ref{eq=28}) at $z=z_{\rm T}$:
  (a)~almost coherent beam,~$\sigma_{_I}=1~\mu$m; (b)~noncoherent beam,~$\sigma_{_I}=0.1~\mu$m.
  Gray vertical strips indicate opaque spaces between slits in the grating $G_{1}$.
  Arrow points out a level of pedestal.
  }
  \label{fig=6}
\end{figure}

 Let us put a detector screen on the distance $L=z_{\rm T}$ from the grating $G_{0}$
 and look out on emergent interference fringes for two cases, namely, for $\sigma_{_I}=1~\mu$m
 and for $\sigma_{_I}=0.1~\mu$m, see~Figs.~\ref{fig=6}(a) and~\ref{fig=6}(b).
 One can see, that at decreasing the dispersion parameter $\sigma_{_I}$ a pedestal supporting the interference fringes emerges.
 Position of the pedestal is pointed out by arrow in Fig.~\ref{fig=6}.
 The pedestal can be found as an absolute minimum for all the interference fringes
\begin{equation}\label{eq=32}
    P_{\rm min} = {\min\limits_{\forall x}}~p(x,z_{_{\rm T}})
\end{equation}
 We can find also an absolute maximum
\begin{equation}\label{eq=33}
    P_{\rm max} = {\max\limits_{\forall x}}~p(x,z_{_{\rm T}})
\end{equation}
 The interferometric visibility be computed by a formula~\citep{CroninMcMorran2006}
\begin{equation}\label{eq=34}
    V = {{P_{\rm max}-P_{\rm min}}\over{P_{\rm max}+P_{\rm min}}}
\end{equation}
 quantifies contrast of the interference fringes.
\begin{figure}[htb!]
  \centering
  \begin{picture}(200,250)(20,10)
      \includegraphics[scale=0.85]{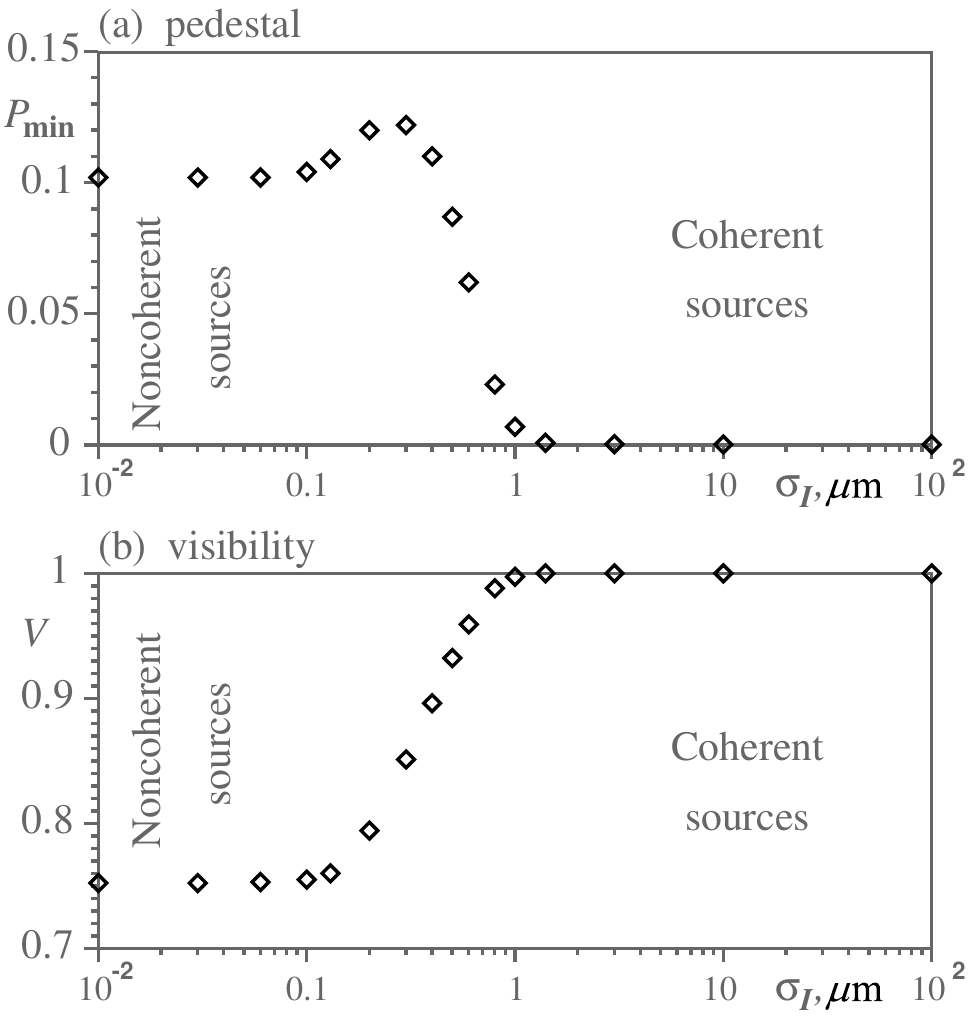}
  \end{picture}
  \caption{
  Pedestal $P_{\rm min}$~(a) and visibility $V$~(b) calculated as functions of the effective coherent width $\sigma_{_I}$ of the beam.
  }
  \label{fig=7}
\end{figure}

 Variations of two parameters, level of the pedestal $P_{\rm min}$ and the visibility $V$, simulated
 as functions of the effective coherent width of the beam $\sigma_{_I}$ ranging from $10^{-2}~\mu$m to $10^{\,2}~\mu$m
 are shown in Figs.~\ref{fig=7}(a) and~\ref{fig=7}(b).
 A crossover is clearly seen within the interval $1~\mu$m down to $0.1~\mu$m.
 It represents a smooth transition from the coherent source to noncoherent
 as the dispersion parameter $\sigma_{_I}$ decreases within the mentioned interval.
\begin{figure*}[htb!]
  \centering
  \begin{picture}(200,380)(170,10)
      \includegraphics[scale=1]{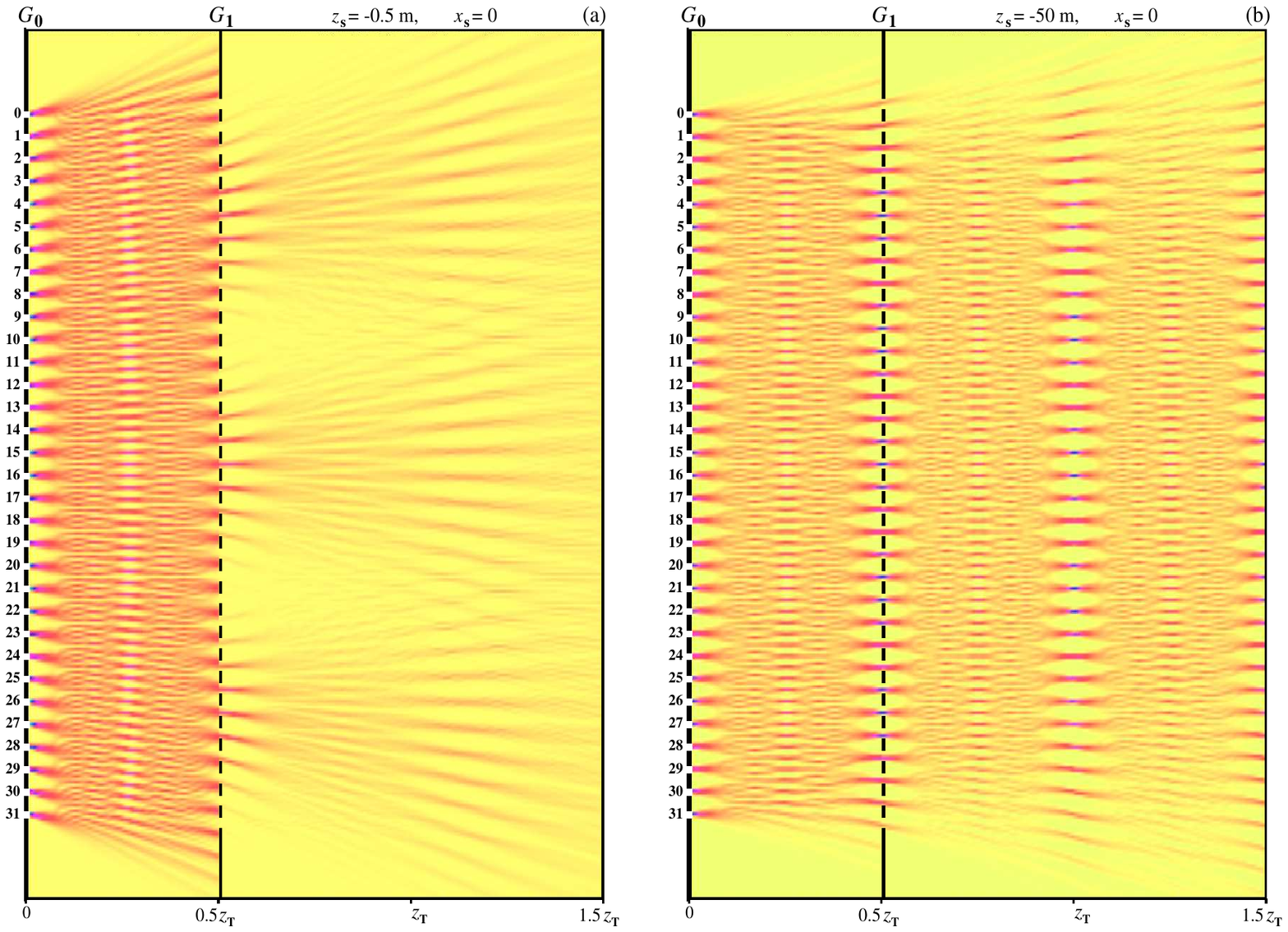}
  \end{picture}
  \caption{
  The density distribution pattern $p(x,z)$: (a)~particle source is spaced closely to the first grating, $z_{\rm s}=-0.5$ m;
  (b)~particle source is spaced far from the first grating, $z_{\rm s}=-50$ m.
  De Broglie wave length $\lambda_{\rm dB}=5$ pm, and $z_{\rm T}=0.1$~m.
  }
  \label{fig=8}
\end{figure*}
\begin{figure}[htb!]
  \centering
  \begin{picture}(200,140)(30,15)
      \includegraphics[scale=0.5]{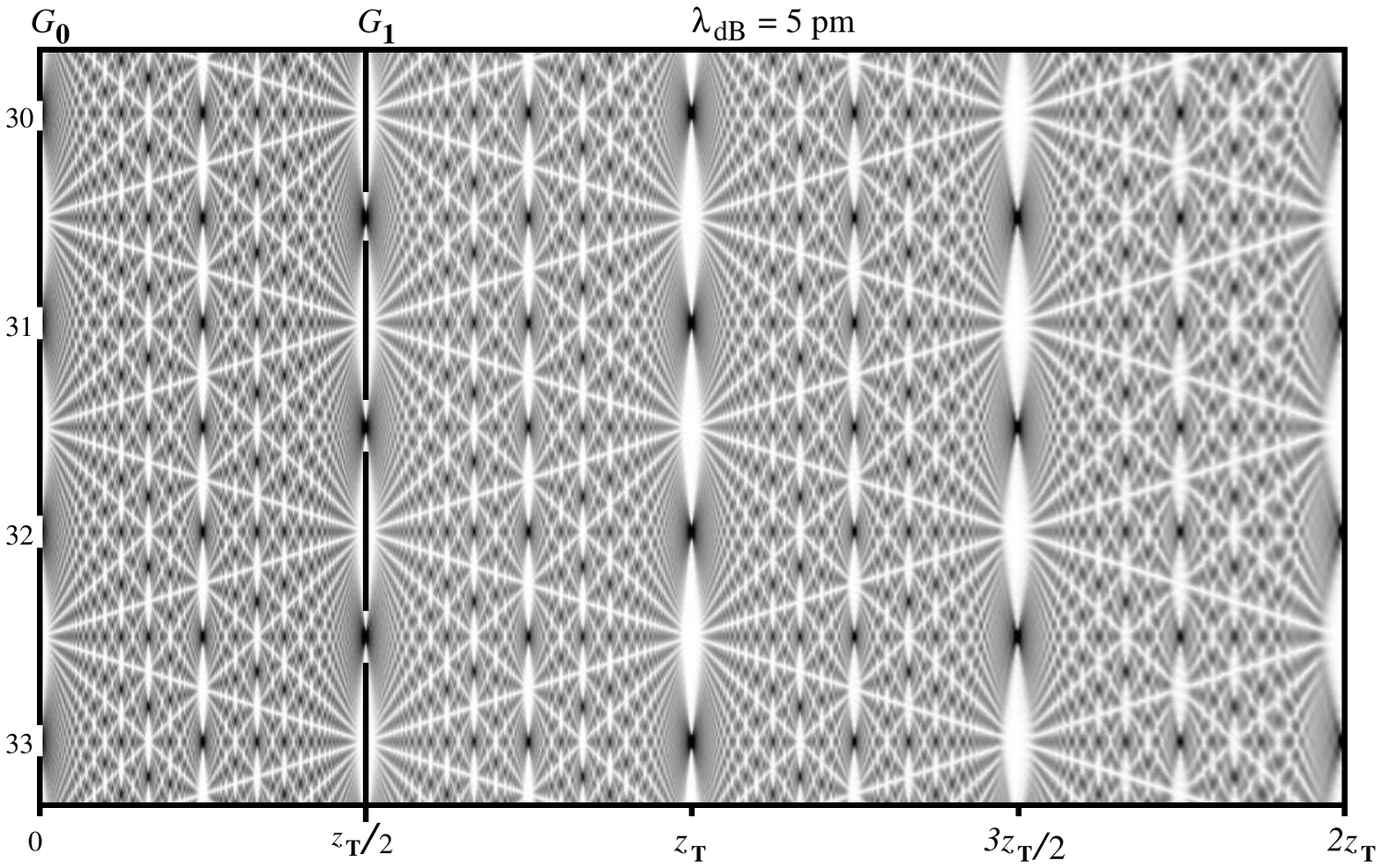}
  \end{picture}
  \caption{
  Talbot carpet on two grating configuration: the gratings $G_{0}$ and $G_{1}$ consists of $N_{0}=64$ and $N_{1}=63$ slits, respectively.
  De Broglie wavelength $\lambda_{\rm dB} = 5$ pm, distance between slits $d=500$ nm, the Talbot length $z_{_{T}}=0.1$ m.
  }
  \label{fig=9}
\end{figure}

 Surprisingly, if radiation from all point sources, all situated near the interferometer, is coherent,
 superposition of these radiations reproduces a perfect interference pattern as if from a single remote source, see Fig.~\ref{fig=5}(a).
 At that, all point sources situated near the grating $G_{0}$ demonstrate radial divergence of rays beyond the grating~$G_{1}$,
 as is seen in Figs.~\ref{fig=4}(a)-(c). To see a general pattern of such a radial divergence we have simulated emergence of
 an interference pattern from two gratings consisting of $N_{0}=32$ and $N_{1}=33$ slits.
 Fig.~\ref{fig=8}(a) shows this interference pattern.
 Because of closely spaced the single source, $z_{\rm s}=-0.5$ m, we see, that there are three groups of divergent rays from
 the second grating. Lateral rays of these groups interfere on a distance of the half Talbot length from the second grating.
 It is remarkable, that superposition of these divergent rays from different coherent sources reproduce the Talbot-like interference pattern
 as shown in Fig.~\ref{fig=5}(a).

 Fig.~\ref{fig=8}(b) demonstrates the very interference but for the point source on $z_{\rm s}=-50$ m distant from the grating $G_{0}$.
 One can believe in this case, that the source has been removed almost to infinity.
 One can see, that Fig.~\ref{fig=8}(b) displays in the near-field the Talbot carpets, that continue behind the second grating.
 For emergence of the perfect Talbot carpet it is necessary to satisfy the following three
 requirements~\citep{Berry1996, BerryKlein1996, BerryEtAl2001}:
 (a)~a~particle beam is paraxial;
 (b)~ratio of de Broglie wavelength, $\lambda_{\rm dB}$, to period of a grating, $d$, tends to zero;
 (c)~number of the slits tends to infinity.
 Fig.~\ref{fig=9} shows the Talbot carpet emergent from the two gratings configuration consisting of $N_{0}=64$ and $N_{1}=63$ slits.
 The Talbot carpet looks the better, the more number of the slits is in the gratings, ideally tending to infinity~\citep{CaseEtAl2009}.

 It is instructive to compare patterns shown in Figs.~\ref{fig=5}(a) and~\ref{fig=9}.
 The first Talbot-like pattern was got by superposing many coherent rays arising from sources situated near the interferometer.
 This pattern is washed out as soon as the rays become noncoherent.
 Whereas the second Talbot pattern is got from a single plane wave incident to the interferometer from infinity.

 Observe that in case of the coherent monochromatic beam,
 presence of the second grating does not affect on emergent the interference pattern.
 That is, the same interference pattern emerges if we would remove the second grating.

 Briefly, the incident beam shows spherical equiphase surfaces of the matter wave when the source is positioned nearby the interferometer.
 At removing the source onto infinity, the spherical equiphase surfaces degenerate into planar equiphase surfaces.
 In that case the spherical wave turns into the plane wave incident on the interferometer.
 The particles momenta are perpendicular to the equiphase surface and particles pass in parallel to the axis of the optical system, the axis $z$.
 Such a particle beam is called the paraxial beam.
 In the next section we obtain this paraxial approximation.

\section{\label{sec:level4}Paraxial approximation}

 The wave functions~(\ref{eq=23})-(\ref{eq=24}) will describe interference in the paraxial approximation
 as soon as a limit $z_{\rm s}\rightarrow-\infty$ will be reached.
 Observe, that in this limit $\Xi_{\,0}=1$, $\Sigma_{\,0,z_{0}}={{\sigma_{0,z_{0}}}/{\sigma_{0,0}}}$ and
\begin{equation}\label{eq=35}
    1-{{1}\over{\Sigma_{\,0,z_{0}}}}={\bf i}\,{{\lambda(z_{1}-z_{0})}\over{4\pi}\sigma_{0,0}\,\sigma_{0,z_{0}}}.
\end{equation}
 For verification see Eqs.~(\ref{eq=17})-(\ref{eq=19}).

 Next, we need also to reinterpret the term
 $D(\Sigma_{\,0,z_{0}},\Sigma_{\,1,z_{1}})\rightarrow D(\sigma_{0,z_{0}},\sigma_{1,z_{1}})$:
\begin{eqnarray}
\nonumber
  &&  D(\sigma_{0,z_{0}}{\textcolor{blue}{,\sigma_{1,z_{1}}}}) = \\
  &&\nonumber \\
  &&
  \sqrt{
    \textcolor{blue}{\Biggl({{z_{}-z_{0}}\over{z_{1}-z_{0}}}\Biggr)
    {{\sigma_{1,z_{1}}}\over{\sigma_{1,0}}}}
    {{\sigma_{0,z_{0}}}\over{\sigma_{0,0}}}
    \textcolor{blue}{- \Biggl({{z_{}-z_{1}}\over{z_{1}-z_{0}}}\Biggr)}
    }.
\label{eq=36}
\end{eqnarray}
 It is instructive to compare this expression with Eq.~(\ref{eq=22}) representing the term $D(\Sigma_{\,0,z_{0}},\Sigma_{\,1,z_{1}})$.

 Here and in the next formula signs colored in blue relate to an area reaching out after the second grating.
 Remaining signs colored in black deal with an area between the gratings.

 As soon as the all reductions have been done, we obtain the wave function in the paraxial approximation,
 i.e., with the source remote onto infinity. It has the following view
\begin{widetext}
\begin{eqnarray}
\nonumber
  \psi(x,z,x_{1},x_{0}) &=& {\displaystyle{{A_{\infty}}\over{D(\sigma_{0,z_{0}}{\textcolor{blue}{,\sigma_{1,z_{1}}}})}}}
   \exp\Biggl\{
   {{\bf i}\pi}\,
    \Biggl[
    \Biggl(
    \textcolor{blue}{{{(x_{}-x_{1})^{2}}\over{\lambda(z_{}-z_{1})}}}
    + {\bf i}\,{{(x_{1}-x_{0})^{2}}\over{4\pi\sigma_{0,0}\sigma_{0,z_{0}}}}
    \Biggr)
    \\
  && \hspace{-14pt}
\textcolor{blue}{-
{{\displaystyle
 {{\lambda(z_{}-z_{1})
 }}
 \over { D(\sigma_{0,z_{0}},\sigma_{1,z_{1}})^{2}
}}
{{\sigma_{0,z_{1}}}\over{\sigma_{0,0}}}
}
\Biggl(
     {{(x_{}-x_{1})}\over{\lambda(z_{}-z_{1})}} -
     {\bf i}\,{{(x_{1}-x_{0})}\over{4\pi\sigma_{0,0}\sigma_{0,z_{0}}}}
\Biggr)^{2}\,}
\Biggr] \Biggr\}.
\label{eq=37}
\end{eqnarray}
\end{widetext}
 Here a factor $A_{\infty}$ replaces $\sqrt{m/(2\pi{\bf i}\hbar\,T)}$. Since at $T\rightarrow\infty$ the quadratic root tends to zero,
 then $A_{\infty}$ tends to zero as well. We will ignore this fact, and suppose let $A_{\infty}$ be some constant.
\begin{figure}[htb!]
  \centering
  \begin{picture}(200,540)(25,15)
      \includegraphics[scale=0.8]{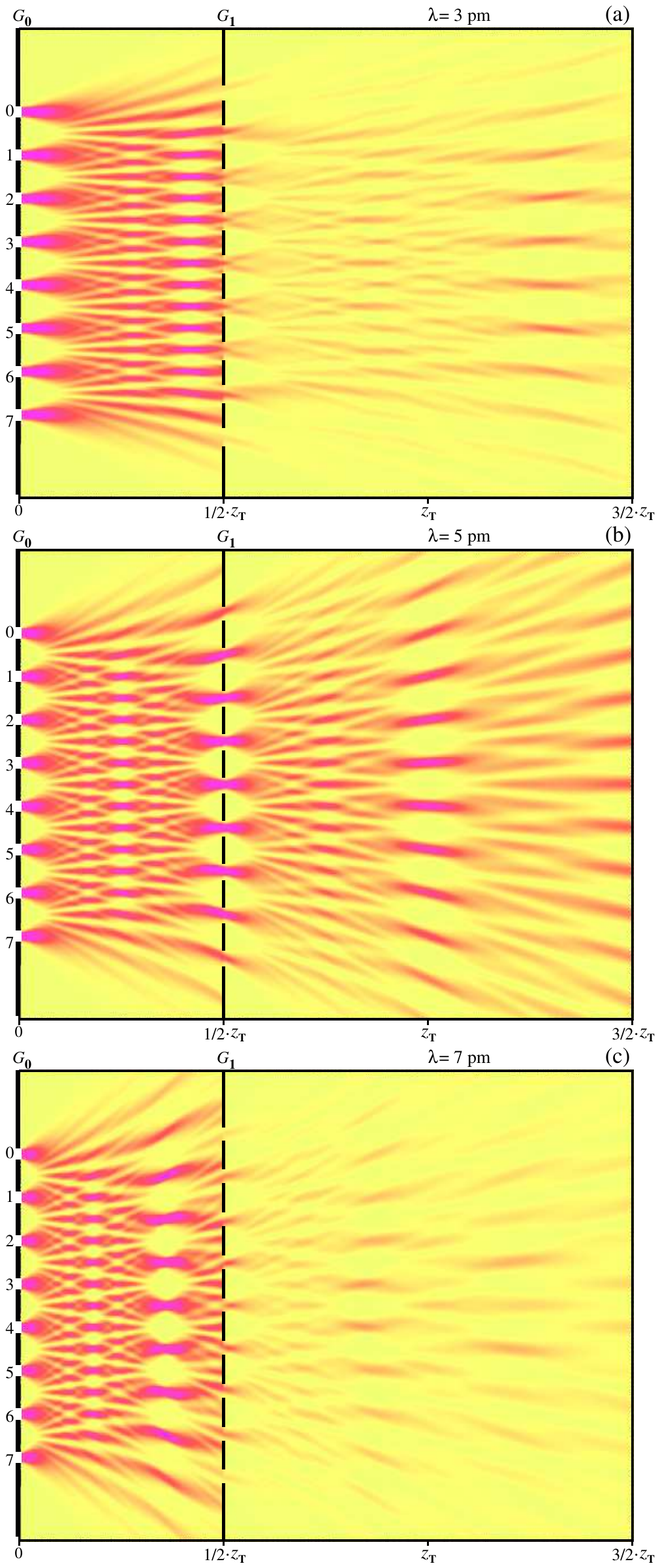}
  \end{picture}
  \caption{
  Density distribution pattern $p(x,z)$ at different de Broglie wavelengths:
  (a)~$\lambda_{_{\rm dB}}=3$~pm, $v_{_{C_{60}}}\approx184$~m/s;
  (b)~$\lambda_{_{\rm dB}}=5$~pm, $v_{_{C_{60}}}\approx110$~m/s;
  (b)~$\lambda_{_{\rm dB}}=7$~pm, $v_{_{C_{60}}}\approx79$~m/s;
  $N_{0}=8$, $N_{1}=9$ and Talbot length $z_{\rm T}=0.1$~m.
  }
  \label{fig=10}
\end{figure}
\begin{figure}
  \centering
  \begin{picture}(200,120)(15,15)
      \includegraphics[scale=0.9]{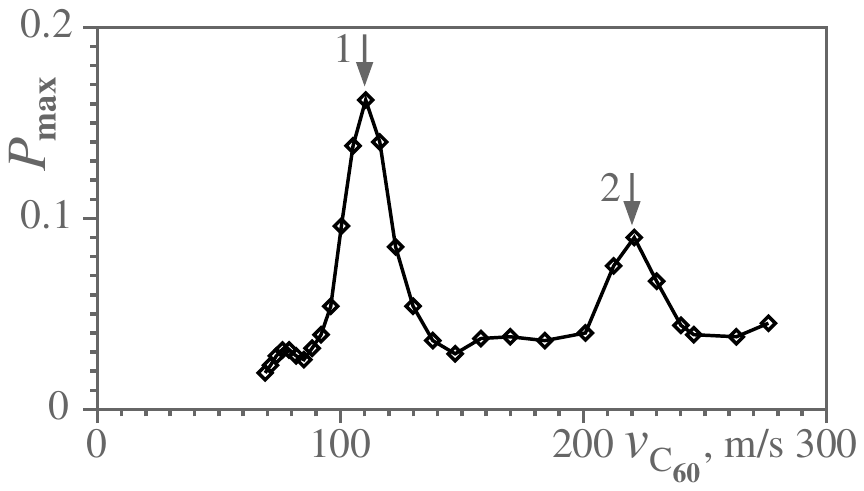}
  \end{picture}
  \caption{
  Emittance $P_{\rm max}$ in the cross-section $z=z_{_{\rm T}}$ vs velocity of the fullerene molecules.
  Arrows 1 and 2 point out to the first and the second resonance harmonics.
  }
  \label{fig=11}
\end{figure}

\begin{figure}
  \centering
  \begin{picture}(200,180)(25,35)
      \includegraphics[scale=0.4]{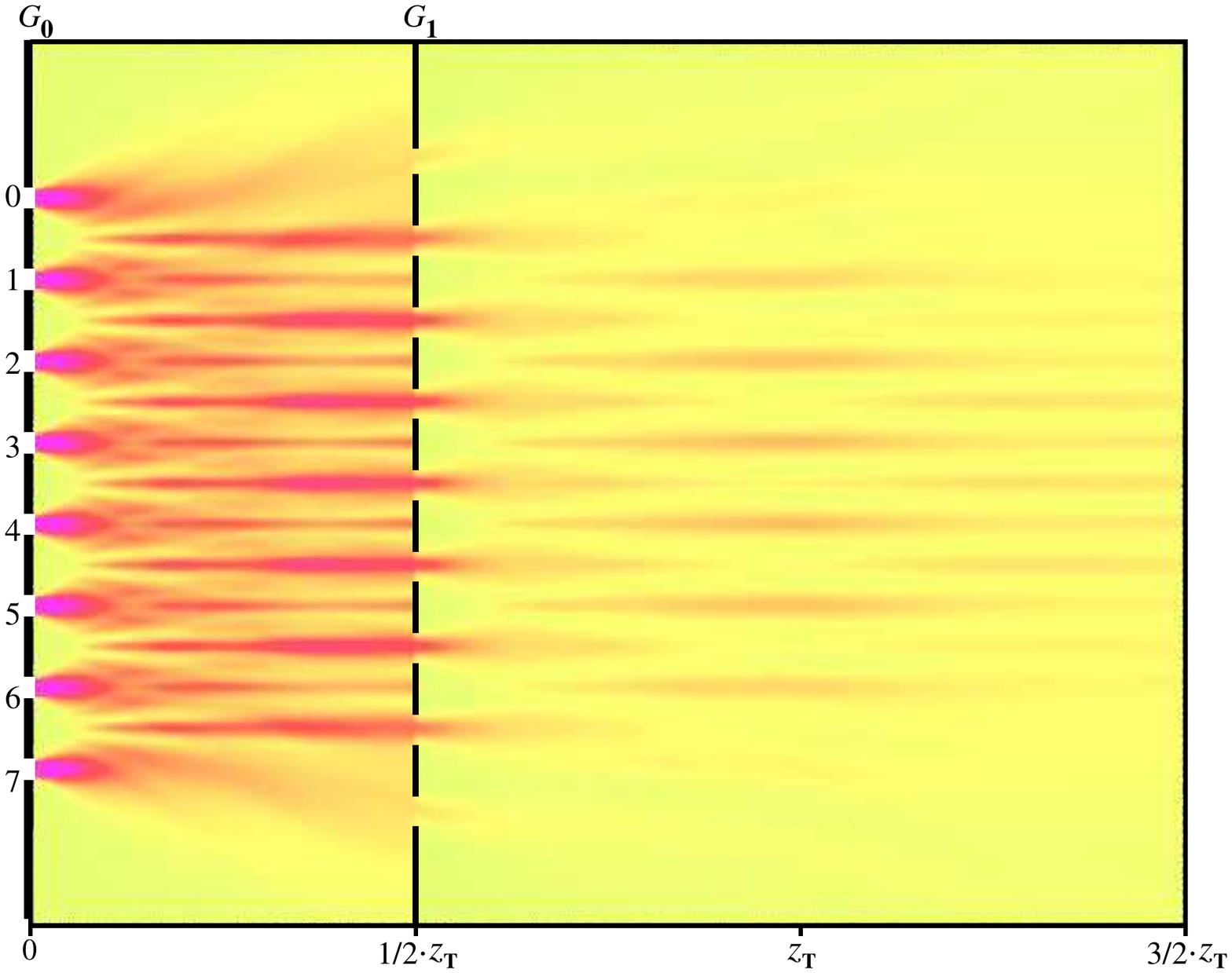}
  \end{picture}
  \caption{
  Density distribution pattern $p(x,z)$ at $N_{0}=8$, $N_{1}=9$
  from a matter wave containing particles
  with different wavelengths from
  $\lambda_{_{\rm dB}}=2$~pm to $\lambda_{_{\rm dB}}=8$~pm.
  }
  \label{fig=12}
\end{figure}

 The wave function~(\ref{eq=37}) describes interference effects emergent both behind the second grating $G_{1}$ and between the gratings $G_{0}$ and $G_{1}$. In order to get the wave pattern between these gratings it is sufficient to set $z_{1}=z$ and $x_{1}=x$. In this case, $\sigma_{1,z_{1}}$ becomes $\sigma_{1,0}$ and $D(\sigma_{0,z_{0}},\sigma_{1,z_{1}})$ reduces to $(\sigma_{0,z_{0}}/\sigma_{0,0})^{1/2}$.
 All terms in~(\ref{eq=37}) containing differences $(x-x_{1})$ and $(z-z_{1})$ disappear.
 To put it bluntly, all terms  in~(\ref{eq=36}) and~(\ref{eq=37}) colored in blue should be removed.

 Let us look on interference patterns emergent on such a device at illumination by matter waves with different wavelengths.
 Figs.~\ref{fig=10}(a) to~\ref{fig=10}(c) show interference patterns for cases of incident particles
 having different de Broglie wavelengths: $\lambda_{\rm dB}=3$~pm,
 $\lambda_{\rm dB}=5$~pm, and $\lambda_{\rm dB}=7$~pm.
 Respectively, velocities of the fullerene molecules at given wavelengths are
 $v_{_{C_{60}}}\approx184$~m/s, $v_{_{C_{60}}}\approx110$~m/s, and $v_{_{C_{60}}}\approx79$~m/s.

 The grating $G_{1}$ is situated at half of the Talbot length.
 Given $\lambda_{\rm dB}=5$~pm and $d=500$~nm the Talbot length is $z_{\rm T}=2d^{\,2}/\lambda_{\rm dB}=0.1$~m.
 Depending on the de Broglie wavelength chosen the interference pattern, emergent between the gratings, discloses different scaling.
 Because of it different interference patterns behind the second grating are formed.
 Most intensive the interference pattern arises at the de Broglie wavelength equal to 5~pm,
 since position of the grating $G_{1}$  has been tuned on the first self-image of $G_{0}$ arising at the same wavelength.

 One can observe a resonance effect at crossing the first self-image of $G_{0}$ by the grating $G_{1}$.
 It can be achieved by changing the wavelength $\lambda_{_{\rm dB}}$
 at crossing the resonance condition $\lambda_{_{\rm dB}}^{\rm res}=2d/z_{_{\rm T}}$.
 In the case under consideration $\lambda_{_{\rm dB}}^{\rm res}=5$~pm.
 Observe, that a maximal emittance from the grating $G_{1}$ is at $\lambda_{_{\rm dB}}=\lambda_{_{\rm dB}}^{\rm res}$,
 when it is positioned exactly on the first self-image of $G_{0}$.
 There can be also high harmonics at
 $n\lambda_{_{\rm dB}}=\lambda_{_{\rm dB}}^{\rm res}$ (here $n$ is integer), when $G_{1}$ is positioned
 on the high order images of $G_{0}$.
 In these cases the emittance from $G_{1}$ quickly drops off with increasing~$n$.
 The emittance to be expressed by a parameter~(\ref{eq=33}) is shown in Fig.~\ref{fig=11}.
 Instead of representing via dependence of  the de Broglie wavelength, here we show the dependence
 via the fullerene velocity $v_{_{C_{60}}}=h/(m_{_{C_{60}}}\lambda_{_{\rm dB}})$. Here $h$ is the Planck constant.
 Evaluations say that for velocities 100--250 m/s
 one needs to support ultralow temperatures ranging about $3\cdot10^{-3}$ to $3\cdot10^{-4}$~K.

 Let us suppose that the remote source emits particles with different de Broglie wavelengths. Distribution over all wavelengths submits
 to the Gaussian with average $\lambda_{\rm dB}=5$~pm and dispersion constant $\sigma_{g}=2.25$ pm. An averaged interference pattern for
 the wavelengths ranged from 3 pm to 8 pm with increment $\delta\lambda_{\rm dB}=0.25$ pm, and under assumption that the sources are noncoherent,
 is shown in Fig.~\ref{fig=12}.
 One can see, that fine-structured details in the interference pattern disappear. They are simply washed out.
 \citet*{ZeilengerEtAl2003} have written for that occasion: "because the detector records the sum of the correspondingly stretched
 or compressed diffraction pictures, the interference pattern would be washed out.
 And in contrast to the spatial contribution, there is no gain in longitudinal (spectral) coherence during free flight."
 Fig.~\ref{fig=12} confirms aforesaid thought.
 Interference patterns disclose equivalent image blur irrespective of choosing of the dispersion constant $\sigma_{g}$.

\section{\label{sec:level5}Gratings with more hard-edged slits}

\begin{figure}
  \centering
  \begin{picture}(200,230)(5,15)
      \includegraphics[scale=0.75]{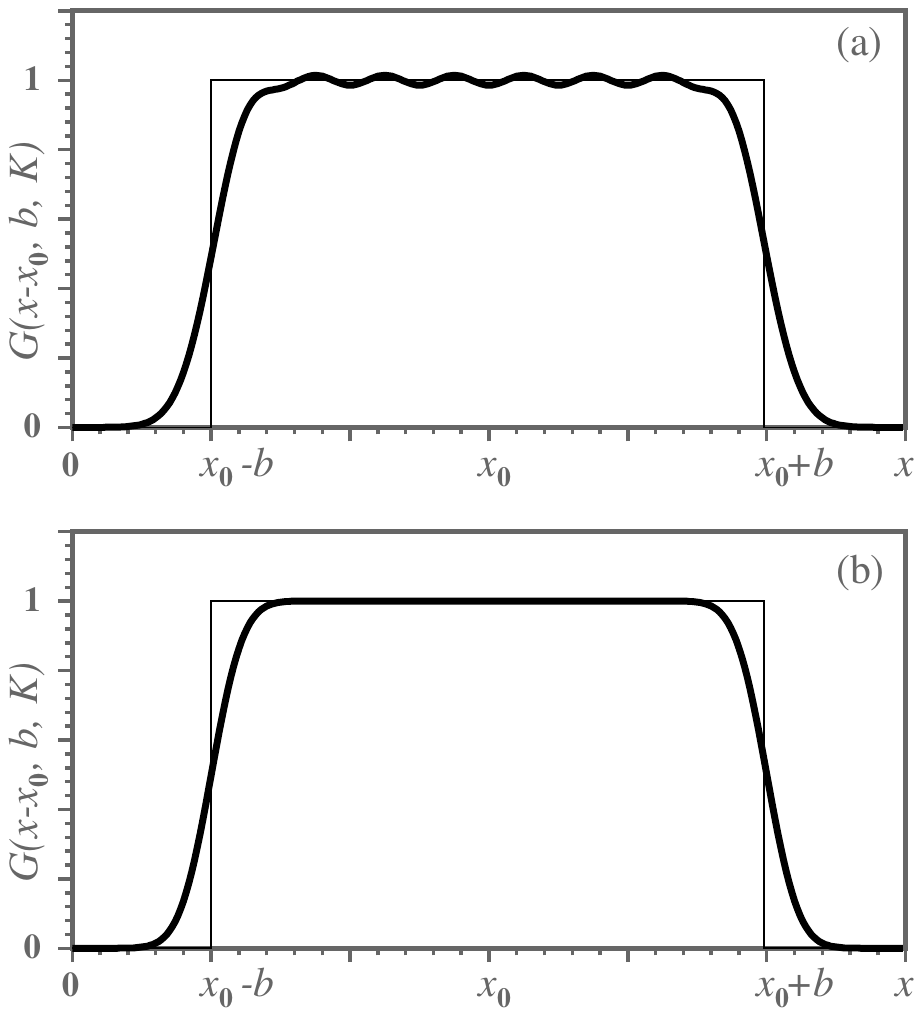}
  \end{picture}
  \caption{
  Approximation of the step function by set of the Gaussian functions presented in~(\ref{eq=38})
  with (a)~$\eta=1$, $K=8$; and (b)~$\eta=1.5$, $K=16$.
  }
  \label{fig=13}
\end{figure}

 Let us compute the path integral~(\ref{eq=4}) for case with more hard-edged slits.
 With that aim we should fill the slits uniformly by a number of the Gaussian functions~(\ref{eq=5}) with more sharp bell curves.
 The step function, that simulate a single slit, can be approximated by the following a set of the Gaussian functions
\begin{eqnarray}
\nonumber
   && G(\xi,b,\eta,K) = \\
   &&
   \hspace{-32pt} {{1}\over{\eta}}\sqrt{{{2}\over{\pi}}}\,\sum\limits_{k=1}^{K}
    \exp\Biggl\{-
    {{\bigl(K\xi-b(K-(2k-1))\bigr)^{2}}\over{2\,(b\,\eta)^{\,2}}}
    \Biggr\}.
\label{eq=38}
\end{eqnarray}
 Here parameter $b$ is a half-width of the slit, real $\eta>0$ is a tuning parameter, and $K$ can take integer values.
 At $K\rightarrow\infty$ this function tends to an infinite collection of the Kronecker deltas,
 that fill everywhere densely the step function.
 Fig.~\ref{fig=13} shows approximation of the step function
 by the set of the Gaussian functions~(\ref{eq=38}) with (a)~$\eta=1$, $K=8$ and (b)~$\eta=1.5$, $K=16$.

 The form-factor $G(\xi_{\,1})$ in Eq.~(\ref{eq=6}) is replaced here by
 the function $G(\xi_{\,1},b_{\,1},\eta_{1},K_{1})$.
 Solution of the integral~(\ref{eq=6}), containing the set of the Gaussian functions~(\ref{eq=38}), satisfies the formula
\begin{eqnarray}
\nonumber
  && {{1}\over{\eta_{\,1}}} \sqrt{{{2}\over{\pi}}}\,\sum\limits_{k=1}^{K_{1}}\;
      \int\limits_{-\infty}^{\infty}
    {\rm e}^{\alpha_{\,k}\,\xi^{2}+\beta_{\,k}\,\xi+\gamma_{\,k}}\,d\xi \\
  &=& {{1}\over{\eta_{\,1}}}\sqrt{{{2}\over{\pi}}}\,\sum\limits_{k=1}^{K_{1}}\;
      \sqrt{{{\pi}\over{-\alpha_{\,k}}}}\,{\rm e}^{-\beta_{\,k}^{2}/4\alpha_{\,k}+\gamma_{\,k}}.
\label{eq=39}
\end{eqnarray}
 Before we compute parameters $\alpha_{\,k}$, $\beta_{\,k}$, $\gamma_{\,k}$
 it is necessary to open the square in Eq.~(\ref{eq=38}).
 After all we find these parameters
 \begin{widetext}
 \begin{enumerate}
   \item the term at $\xi_{1}^{2}$:
\begin{equation}\label{eq=40}
   \alpha_{\,k} = {{{\bf i}m}\over{2\hbar}}
    \Biggl(
    {{1}\over{\tau_{1}}} + {{1}\over{\tau_{0}}}
    + {\bf i}\,{{\hbar}\over{m\eta_{1}^{\,2}}}{{K_{1}^{2}}\over{b_{\,1}^{\,2}}}
    -{{1}\over{\tau_{0}^{2}(1/\tau_{0}+1/T+{\bf i}\hbar/mb_{0}^{2})}}
    \Biggr);
\end{equation}
   \item the term at $\xi_{1}$:
\begin{eqnarray}
\nonumber
   \beta_{\,k} &=& -2\,{{{\bf i}m}\over{2\hbar}} \Biggl(
     {{(x_{2}-x_{1})}\over{\tau_{1}}} - {{(x_{1}-x_{0})}\over{\tau_{0}}}
     + {{(x_{1}-x_{0})/\tau_{0}^{2}-(x_{0}-x_{\rm s})/\tau_{0}T}\over
     {(1/\tau_{0}+1/T+{\bf i}\hbar/m\,b_{0}^{2})}} \\
  &+& {\bf i}\,{{\hbar}\over{m\eta_{1}^{\,2}}} {{K_{1}}\over{b_{\,1}}} (K_{1}-(2k-1))
    \Biggr);
\label{eq=41}
\end{eqnarray}
   \item the term free from $\xi_{1}$:
\begin{eqnarray}
\nonumber
    \gamma_{\,k} &=& {{{\bf i}m}\over{2\hbar}} \Biggl(
    {{(x_{2}-x_{1})^{2}}\over{\tau_{1}}} + {{(x_{1}-x_{0})^{2}}\over{\tau_{0}}} + {{(x_{0}-x_{\rm s})^{2}}\over{T}}
    - {{((x_{1}-x_{0})/\tau_{0}-(x_{0}-x_{\rm s})/T)^{2}}\over{(1/\tau_{0}+1/T+{\bf i}\hbar/mb_{0}^{2})}} \\
    &+& {\bf i}\,{{\hbar}\over{m\eta_{1}^{\,2}}}\bigl(K_{1}-(2k-1)\bigr)^{2}
    \Biggr).
\label{eq=42}
\end{eqnarray}
 \end{enumerate}
\end{widetext}

 It is instructive to compare these parameters with those presented in Eqs.~(\ref{eq=8}), ~(\ref{eq=9}), and~(\ref{eq=10}).
 Differences easily strike the eye.
 The amplitude factor~(\ref{eq=12}) for our new task reads
\begin{equation}\label{eq=43}
    A = {{1}\over{\eta_{\,1}}}\sqrt{{{2}\over{\pi}}}\cdot
    \sqrt{{{m}\over{{\bf i}\,2\pi\,\hbar T}}}\cdot{{1}\over{D(b_{0},{b}_{1})}}.
\end{equation}
 A minor difference of the amplitude factors~(\ref{eq=12}) and~(\ref{eq=43}) is conditioned by
 an additional factor $(2/\pi)^{1/2}/\eta_{\,1}$ presented in Eq.~(\ref{eq=39}).
 In particular, at $\eta_{1}\approx1.5$ and $K_{1}=1$ the amplitude factor~(\ref{eq=43}) will be twice as little  of the factor~(\ref{eq=12}).
 The denominator $D(b_{0},{b}_{1})$ here has the following view
\begin{widetext}
\begin{equation}\label{eq=44}
    D(b_{0},{b}_{1}) =
     \sqrt{
 \displaystyle
  \Biggl(1+{{\tau_{1}}\over{\tau_{0}}}\Biggr)
  \Biggl({\displaystyle
  1 + {{{\bf i}K_{1}^{\,2}\hbar\tau_{1}}\over{m\,\eta_{\,1}^{\,2}\,{b}_{\,1}^{\,2}(1+\tau_{1}/\tau_{0})}}
  }\Biggr)
  \Biggl(1+{{\tau_{0}}\over{T}}\Biggr)
  \Biggl({\displaystyle
  1 + {{{\bf i}\hbar\tau_{0}}\over{m\,b_{0}^{\,2}(1+\tau_{0}/T)}}
  }\Biggr)
  - {{\tau_{1}}\over{\tau_{0}}}}.
\end{equation}
 In turn, the therm $\gamma_{\,k}-\beta_{\,k}^{2}/4\alpha_{\,k}$ reads
\begin{eqnarray}
\nonumber 
 &&  \gamma_{\,k}-\beta_{\,k}^{2}/4\alpha_{\,k} =
  {{{\bf i}m}\over{2\hbar}} 
  \Biggl[
  \Biggl(
    {{(x_{2}-x_{1})^{2}}\over{\tau_{1}}} + {{(x_{1}-x_{0})^{2}}\over{\tau_{0}}} + {{(x_{0}-x_{\rm s})^{2}}\over{T}}
    - {{((x_{1}-x_{0})/\tau_{0}-(x_{0}-x_{\rm s})/T)^{2}}\over{((\tau_{0}+T)/T\tau_{0}+{\bf i}\hbar/m\,b_{0}^{2})}}
     \\
\nonumber
 && \hspace{96pt} +\; {\bf i}\,{{\hbar}\over{m\,\eta_{\,1}^{\,2}}}\bigl(K_{1}-(2k-1)\bigr)^{2} \Biggr) \\
 &&
   -\;{{\displaystyle\Biggl({{(x_{2}-x_{1})}\over{\tau_{1}}} - {{(x_{1}-x_{0})}\over{\tau_{0}}}
     + {{(x_{1}-x_{0})/\tau_{0}-(x_{0}-x_{\rm s})/T}\over
     {\tau_{0}((\tau_{0}+T)/T\tau_{0}+{\bf i}\hbar/m\,b_{0}^{\,2})}}
    + {\bf i}\,{{\hbar}\over{m\,\eta_{\,1}^{\,2}}} {{K_{1}}\over{{b}_{\,1}}} (K_{1}-(2k-1))
    \Biggr)^{2}}
  \over{\displaystyle
 \Biggl(
  {\displaystyle{{\tau_{0}+T}\over{T\tau_{0}}} + {{{\bf i}\hbar}\over{m\,b_{0}^{2}}}}
 \Biggr)^{-1}
{{1}\over{\tau_{0}\tau_{1}D(b_{0},{b}_{1})}}
 }}
    \hspace{-6pt}
    \left. \matrix{\cr\cr\cr\cr\cr}\right]
\label{eq=45}
\end{eqnarray}
\end{widetext}

\subsection{\label{subsec:level5A}Series of replacements}

 In order to execute series of the replacements as in Subsec.~\ref{subsec:level2A}
 it is proposed to introduce the following scaling change for the  half-width $b_{1}$
\begin{equation}\label{eq=46}
    {{\eta_{\,1}b_{\,1}}\over{K_{1}}}\rightarrow b_{\,1}^{\,'}.
\end{equation}
 And consequently, we will have in mind that the parameters $\sigma_{1,0}$, $\sigma_{1,\tau_{1}}$, $\Sigma_{1,z_{1}}$
 defined in Eqs.~(\ref{eq=14}),~(\ref{eq=15}),~\ref{eq=18}) can contain the very terms $K_{1}$ and $\eta_{\,1}$.
 After all, the parameters $\Sigma_{0,z_{0}}$  and $\Sigma_{1,z_{1}}$ in this task is rewritten as follows
\begin{eqnarray}
\label{eq=47}
  \Sigma_{0,z_{0}} &=& {{z_{1}-z_{\rm s}}\over{z_{0}-z_{\rm s}}}
   + {\bf i} {{\lambda(z_{1}-z_{0})}\over{4\pi\sigma_{0,0}^{\,2}}}, \\
  \Sigma_{1,z_{1}} &=& {{z_{2}-z_{0}}\over{z_{1}-z_{0}}}
   + {\bf i} {{\lambda(z_{2}-z_{1})}\over{4\pi\sigma_{1,0}^{\,2}}}{{K_{1}^{\,2}}\over{\eta_{\,1}^{\,2}}}.
\label{eq=48}
\end{eqnarray}
 Here $\sigma_{0,0}$ and $\sigma_{1,0}$ have the original forms presented in~(\ref{eq=14}).

 The phase term $\gamma_{\,k}-\beta_{\,k}^{2}/4\alpha_{\,k}$, that has a rather complex form~(\ref{eq=45}), reads
\begin{widetext}
\begin{eqnarray}
\nonumber
 &&  \gamma_{\,k}-\beta_{\,k}^{2}/4\alpha_{\,k} = {{\bf i}\pi}\,
    \Biggl[
    \Biggl(
    {{(x_{2}-x_{1})^{2}}\over{\lambda(z_{2}-z_{1})}}
    + {{(x_{1}-x_{0})^{2}}\over{\lambda(z_{1}-z_{0})}}
   \Biggl(
    1 -
    {{\Xi_{\,0}^{\,2}}\over{\Sigma_{0,z_{0}}}}
    \Biggr)
    + {{(x_{0}-x_{\rm s})^{2}}\over{\lambda(z_{0}-z_{\rm s})}}
 +\; {\bf i}\,{{\bigl(K_{1}-(2k-1)\bigr)^{2}}\over{2\pi\,\eta_{\,1}^{\,2}}}
    \Biggr)
    \\
  && \hspace{-14pt}
-
{{\displaystyle
 {{\lambda(z_{2}-z_{1})
 \Sigma_{0,z_{0}}
 }}
 \over { D(\Sigma_{0,z_{0}},\Sigma_{1,z_{1}})^{2}
}}
}
\Biggl(
     {{(x_{2}-x_{1})}\over{\lambda(z_{2}-z_{1})}} -
     {{(x_{1}-x_{0})}\over{\lambda(z_{1}-z_{0})}}
\Biggl(
     1 -
 {{\Xi_{\,0}}\over{\Sigma_{0,z_{0}}}}
\Biggr) + {\bf i}\,{{K_{1}(K_{1}-(2k-1))}\over{2\pi\,\eta_{\,1}^{\,2}\,{b}_{1}}}
\Biggr)^{2}\,
\Biggr],
\label{eq=49}
\end{eqnarray}
 Here $D(\Sigma_{0,z_{0}},\Sigma_{1,z_{1}})$ is represented in~(\ref{eq=22}), but with $\Sigma_{1,z_{1}}$ to be loaded from~(\ref{eq=48}).

 Now we can write the wave functions describing appearance of a particle both between the gratings and behind the second grating.

\subsection{\label{subsec:level5B}Matter waves behind the gratings $G_{0}$ and $G_{1}$}

 Evolving a particle from a single slit in $G_{0}$ to $G_{1}$ is described by the very wave function~(\ref{eq=24}):
\begin{equation}\label{eq=50}
    \psi(x,z,x_{0},x_{\rm s}) =
    {\displaystyle{\sqrt{\displaystyle{{m}\over{{\bf i}2\pi^{}\hbar T\Sigma_{0,z_{0}}}}}}}\cdot
   \exp\Biggl\{
   {{\bf i}\pi}\,
    \Biggl[
     {{(x_{}-x_{0})^{2}}\over{\lambda(z_{}-z_{0})}}
   \Biggl(
    1 -
    {{\Xi_{\,0}^{\,2}}\over{\Sigma_{0,z_{0}}}}
    \Biggr)
    + {{(x_{0}-x_{\rm s})^{2}}\over{\lambda(z_{0}-z_{\rm s})}}
    \Biggr] \Biggr\}.
\end{equation}
 Observe that equivalence of the wave functions~(\ref{eq=50}) and~\ref{eq=24}) is due to the fact
 that we use the same grating $G_{0}$ in the both cases.
 But behind the grating $G_{1}$, having more hard-edged slits,
 movement of the particle from a single slit in $G_{1}$ is described by a wave function
\begin{eqnarray}
\nonumber
  &&\hspace{64pt} \psi(x,z,x_{1},x_{0},x_{\rm s},K_{1}) = \displaystyle {{1}\over{\eta_{\,1}}} {\sqrt{{{2}\over{\pi}}}}\cdot{\displaystyle{
  {\sqrt{\displaystyle{{m}\over{{\bf i}2\pi^{}\hbar T}}}}\over{D(\Sigma_{0,z_{0}},\Sigma_{1,z_{1}})}}}
  \sum\limits_{k=1}^{K_{1}}\\
\nonumber
 && \exp\Biggl\{
     \Biggl[
    \Biggl(
    {{(x_{}-x_{1})^{2}}\over{\lambda(z_{}-z_{1})}}
    + {{(x_{1}-x_{0})^{2}}\over{\lambda(z_{1}-z_{0})}}
   \Biggl(
    1 -
    {{\Xi_{\,0}^{\,2}}\over{\Sigma_{0,z_{0}}}}
    \Biggr)
    + {{(x_{0}-x_{\rm s})^{2}}\over{\lambda(z_{0}-z_{\rm s})}}
 +\; {\bf i}\,{{\bigl(K_{1}-(2k-1)\bigr)^{2}}\over{2\pi\,\eta_{\,1}^{\,2}}}
    \Biggr)
    \\
  && \hspace{-14pt}
  - {{\displaystyle
 {{\lambda(z_{}-z_{1})
 \Sigma_{0,z_{0}}
 }}
 \over { D(\Sigma_{0,z_{0}},\Sigma_{1,z_{1}})^{2}
}}
}
\Biggl(
     {{(x_{}-x_{1})}\over{\lambda(z_{}-z_{1})}} -
     {{(x_{1}-x_{0})}\over{\lambda(z_{1}-z_{0})}}
\Biggl(
     1 -
 {{\Xi_{\,0}}\over{\Sigma_{0,z_{0}}}}
\Biggr) + {\bf i}\,{{K_{1}(K_{1}-(2k-1))}\over{2\pi\,\eta_{\,1}^{\,2}\,{b}_{1}}}
\Biggr)^{2}\,
\Biggr]
  \Biggr\}.
\label{eq=51}
\end{eqnarray}
\end{widetext}
 One can see, that with $K_{1}=1$ the wave function~(\ref{eq=51}), accurate to the factor $(2/\pi)^{1/2}/\eta_{\,1}$, is simply the function~(\ref{eq=23}).

 Interference patterns arise as superpositions of the above wave functions radiated from all slits of the grating $G_{0}$ and from all slits
 of the grating $G_{1}$.
 These wave functions of the matter waves emitted from all slits of the gratings $G_{0}$ and $G_{1}$
 have been written out in Eqs.~(\ref{eq=25}) and~(\ref{eq=26}).
 Note only, that the wave function $|\Psi_{1}\rangle$ contains extra parameters, integer $K_{1}$ and real $\eta_{\,1}$.
\begin{figure}[htb!]
  \centering
  \begin{picture}(200,540)(25,15)
      \includegraphics[scale=0.8]{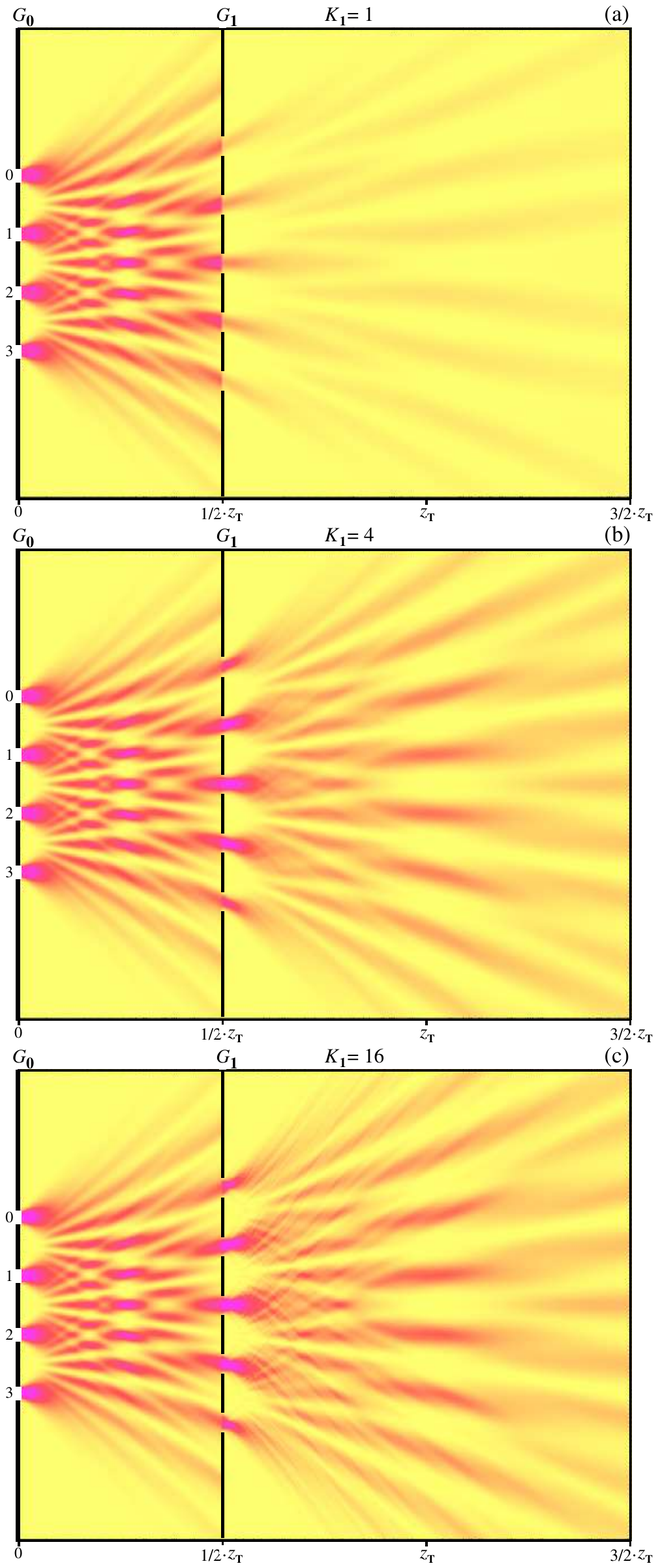}
  \end{picture}
  \caption{
  Density distribution pattern $p(x,z)$
  in the near-field region $z\in(0,1.5 z_{\rm T})=(0,0.15)$ m, $N_{0}=4$, $N_{1}=5$,
  de Broglie wavelength $\lambda_{\rm dB}=5$ pm, Talbot length $z_{\rm T}=0.1$~m.
  Distance from $G_{0}$ to the source is $z_{\rm s}=-0.5$ m, $x_{\rm s}=0~\mu$m :
  (a)~$K_{1}=1$;
  (b)~$K_{1}=4$;
  (c)~$K_{1}=16$.
  For all cases $\eta_{\,1}=1.5$.
  }
  \label{fig=14}
\end{figure}
\begin{figure}[htb!]
  \centering
  \begin{picture}(200,300)(38,15)
      \includegraphics[scale=0.45]{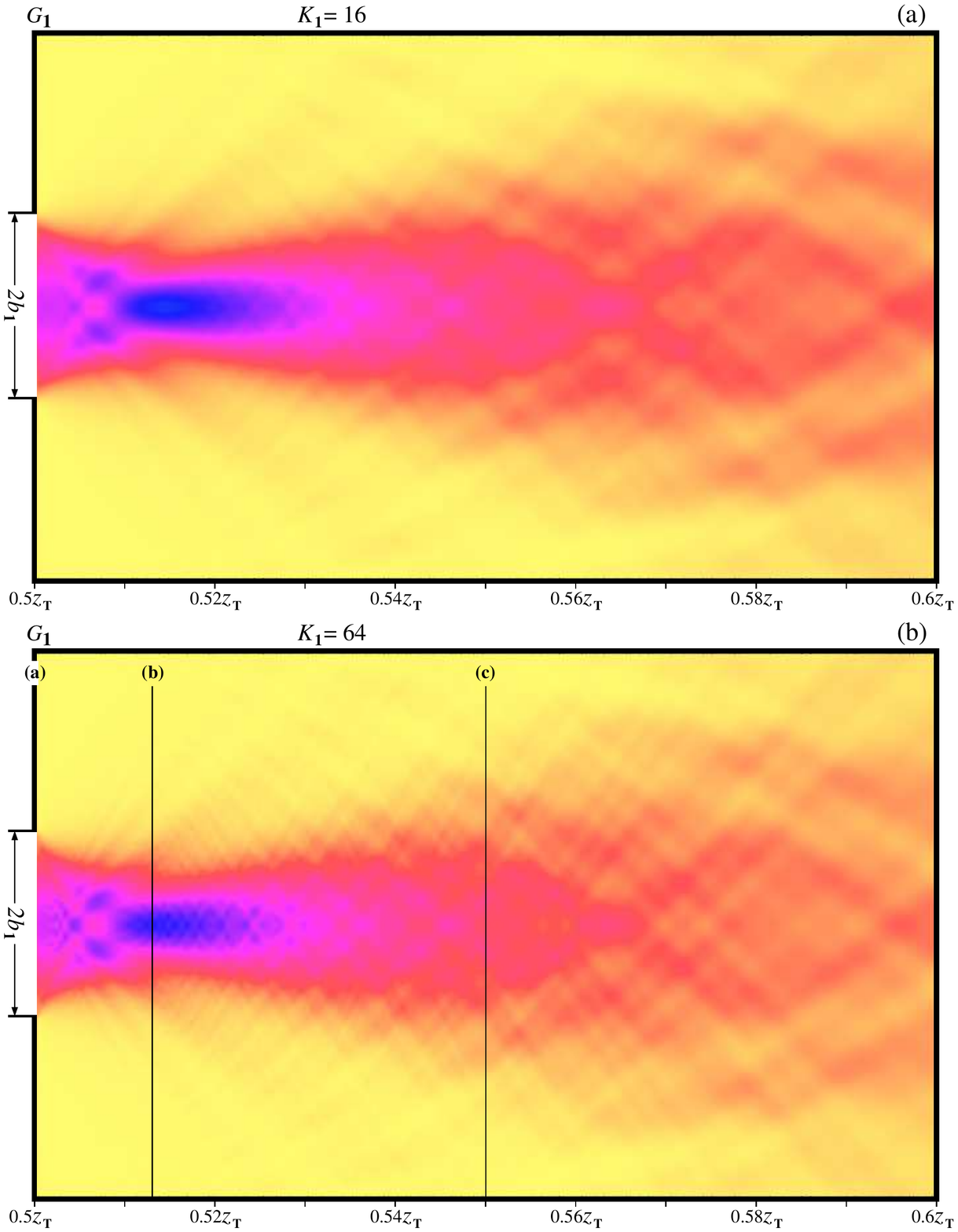}
  \end{picture}
  \caption{
  Cutting out radiation from the central slit of $G_{1}$ shown in Fig.~\ref{fig=14}(c): (a)~reproduced for case of~$K_{1}=16$;
  (b)~reproduced for case of~$K_{1}=64$.
  Blue patches near the slit are most higher intensities of the radiation.
  }
  \label{fig=15}
\end{figure}
\begin{figure}
  \centering
  \begin{picture}(200,100)(15,15)
      \includegraphics[scale=0.8]{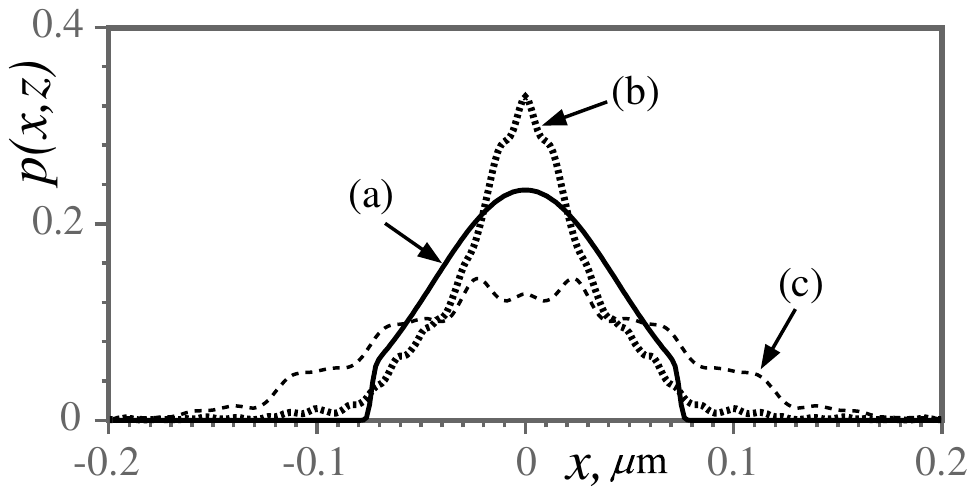}
  \end{picture}
  \caption{
 Profiles of the probability density $p(x,z)$ within cross-sections
 (a)~$z=0.5z_{_{\rm T}}$; (b)~$z=0.513z_{_{\rm T}}$; (c)~$z=0.55z_{_{\rm T}}$
 of the interference pattern shown in Fig.~\ref{fig=15}.
  }
  \label{fig=16}
\end{figure}

 Figs.~\ref{fig=14}(a),~\ref{fig=14}(b), and~\ref{fig=14}(c) show interference patterns emergent behind gratings containing
 only $N_{0}=4$ and $N_{1}=5$ slits. Slits of the grating $G_{1}$ are simulated by the set of the Gaussian functions~(\ref{eq=38})
 with the parameter $\eta_{\,1}$ equal to $1.5$ and for different $K_{1}$: (a)~$K_{1}=1$;
 (b)~$K_{1}=4$; and (c)~$K_{1}=16$.
 More rapid divergence of the rays near the slits is seen to arise in the case of large $K_{1}$.
 These divergent rays form finely ruled interference fringes arising
 prior to the first Talbot length $z_{_{\rm T}}$, see Fig.~\ref{fig=14}(c).

 Observe, that behind double the Talbot length, $2z_{_{\rm T}}$, the interference fringes become equivalent
 independently of choosing $K_{1}$. Qualitative difference of intensities seen in Fig.~\ref{fig=14}(a), on the one hand,
 and in Figs.~\ref{fig=14}(b)-\ref{fig=14}(c), on the other hand, is due to presence of the factor $(2/\pi)^{1/2}/\eta_{\,1}\approx0.5$.
 This factor provides a good coincidence of a top plateau of the function~(\ref{eq=38}) with that of the step function for $K_{\,1}>1$.
 Whereas in case of $K_{1}=1$, height of the Gaussian function becomes  twice as little.
 In particular, the intensity in Fig.~\ref{fig=14}(a) is also twice as little than that in Fig.~\ref{fig=4}(a) for the same reason.

 One can observe a surprising phenomenon nearby the slits of the grating $G_{1}$ as $K_{1}$ increases -- the larger $K_{1}$ the better.
 The beam outgoing from a slit, first, goes through focusing. After it passes a smallest diameter
 (it overcomes so called the beam waist~\citep{CroninMcMorran2006}) the beam begins to diverge.
 A detail pattern of this phenomenon in the vicinity of the central slit of $G_{1}$, see Fig.~\ref{fig=14}(c),
 is demonstrated in Fig.~\ref{fig=15}(a).
 The outgoing intensity pattern shown in a range from $0.5z_{_{\rm T}}$ to $0.6z_{_{\rm T}}$ is seen to have a complex organization.
 A tongue-like "jet" outgoing from the slit, first, undergoes squeezing. Dark patches are places where the intensity reaches maximal values.
 The most maximal value is reached where the "jet" goes through the beam waist. Well organized small dark patches seen in the figure precede
 the most dark central patch. These small dark patches originate from center of the slit.
 And as they approach towards the most dark central patch, weak divergent rays can be seen are radiated far aside.
 More detailed jet outgoing pattern simulated for $K_{1}=64$ is shown in Fig.~\ref{fig=15}(b).

 Fig.~\ref{fig=16} shows some profiles of the probability density $p(x,z)$ taken from the interference pattern,
 that is obtained at given $K_{1}=64$ and $\eta_{1}=1.5$, Fig.~\ref{fig=15}(b).
 Integrals along the profiles give almost equal values.
 Insignificant discrepancies are conditioned by dissipation of the matter wave beyond an integration interval.

 The profiles (a), (b), and (c) depicted in Fig.~\ref{fig=16} are captured from cross-sections,
 that are marked by lines (a), (b), and (c) in Fig.~\ref{fig=15}(b).
 They are at~$z=0.5z_{_{\rm T}}$,~$z=0.513z_{_{\rm T}}$, and~$z=0.55z_{_{\rm T}}$, respectively.
 The profile (a) abuts on the slit screen.
 We can see in Fig.~\ref{fig=16} that the profile~(b)  clearly demonstrates against the background of the profile (a)
 a squeezing of the beam together with increasing its intensity along the center. Next, the beam begins to disperse
 as distance from the slit increases, see, for example, the profile (c) in the same figure.

\subsection{\label{subsec:level5C}Emergence of the focusing spot}

 Emergence of the focusing spot with increasing number of the Gaussian functions in~(\ref{eq=38}), i.e., with increasing $K_{1}$ is an astonishing phenomenon at shaping slits with more hard edges.
 An effect of squeezing beam right after the slit can be understood at choosing of the slits that are approximated
 by the curve~(\ref{eq=38}) with small $K_{1}$ and $\eta_{\,1}$.
 Let us study such an effect (a) with increasing $K_{1}$ at fixed $\eta_{1}$; and (b) with increasing $\eta_{1}$ at fixed $K_{1}$.

\begin{figure}[htb!]
  \centering
  \begin{picture}(200,150)(20,5)
      \includegraphics[scale=0.5]{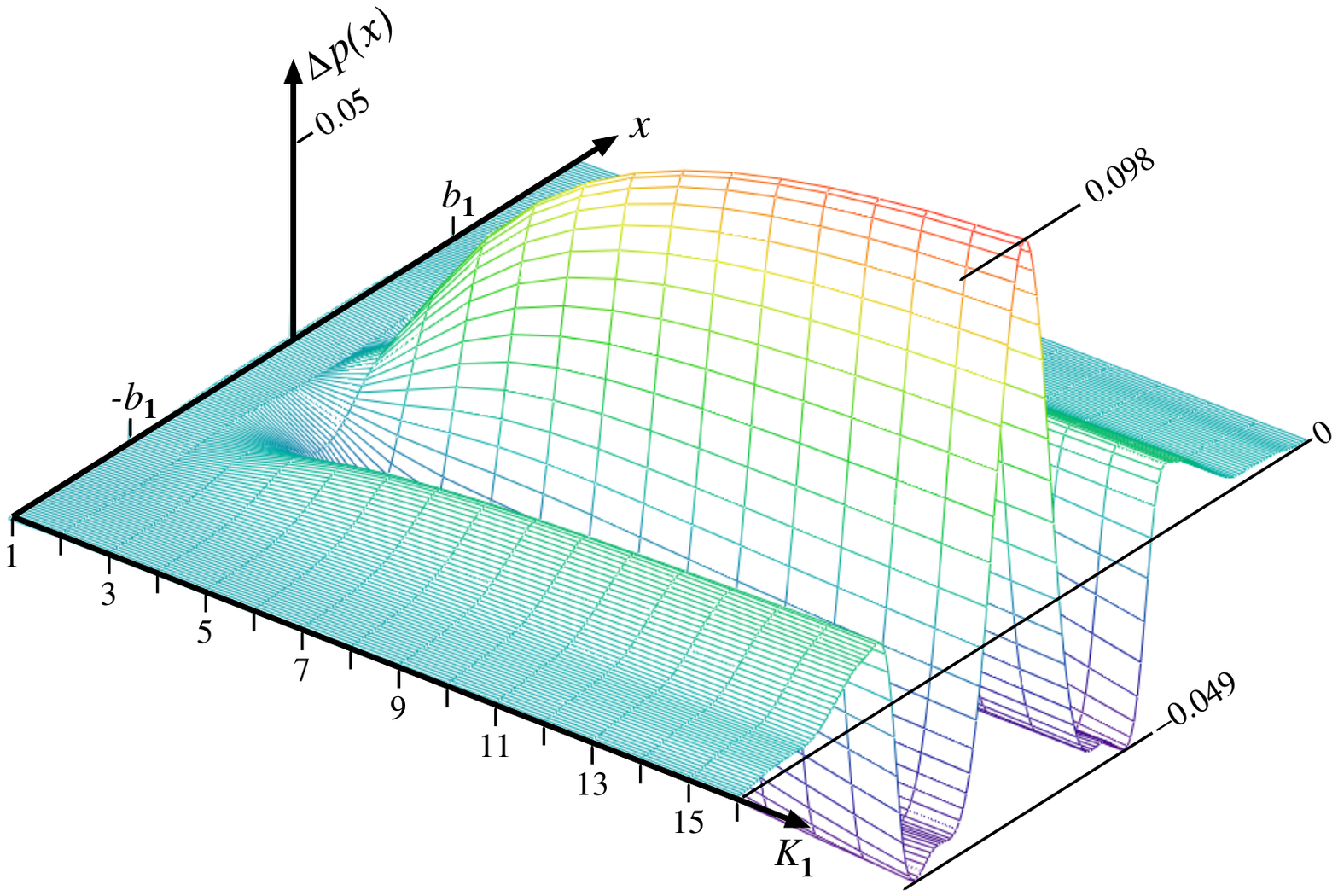}
  \end{picture}
  \caption{
  Difference $\Delta p(x)$ of  two profiles of the probability density $p(x,z)$
  given on the slit ($z_{a}=0.5z_{_{\rm T}}$) and in the cross-section of the beam waist ($z_{\,b}=0.513z_{_{\rm T}}$)
  as a function of $K_{1}$.
  The real parameter $\eta_{1}$ is equal to $1.5$.
  }
  \label{fig=17}
\end{figure}

 Let us consider, first, emergence of the focusing spot with increasing integer $K_{1}$ from 1 to 16 at fixed $\eta_{1}=1.5$.
 For this aim we propose to evaluate a difference of  two profiles of the probability density $p(x,z)$
 given on the slit ($z_{a}=0.5z_{_{\rm T}}$) and in the cross-section of the beam waist ($z_{\,b}=0.513z_{_{\rm T}}$):
\begin{equation}\label{eq=52}
    \Delta p(x) =
    p(x,z_{\,b}) - p(x,z_{a}).
\end{equation}
 As seen in Fig.~\ref{fig=16} this difference can have a large positive hill in center of the beam bounded by  wells from the both sides.
 All lies within a space of the slit.
 Fig.~\ref{fig=17} shows a series of the differences $\Delta p(x)$ to be calculated for $K_{1}$ ranging from 1 to 16 and at fixed $\eta_{1}=1.5$.
 One can see, given $K_{1}$ equal to 1 there is no $\Delta p(x)$ different from zero. As $K_{1}$ increases from 1 to 16 $\Delta p(x)$ grows quickly enough
 and at $K_{1}=16$ its peak reaches almost~$0.1$.
 It is the focusing spot bounded by the wells from the both sides.
 A depth of the wells is about $-0.05$ and they lie between edges of the slit, between $-b_{1}$ and~$b_{1}$.
 The difference~(\ref{eq=52}) brings to light the focusing effect of outgoing from the slit the beam well enough.

\begin{figure}[htb!]
  \centering
  \begin{picture}(200,310)(5,10)
      \includegraphics[scale=0.75]{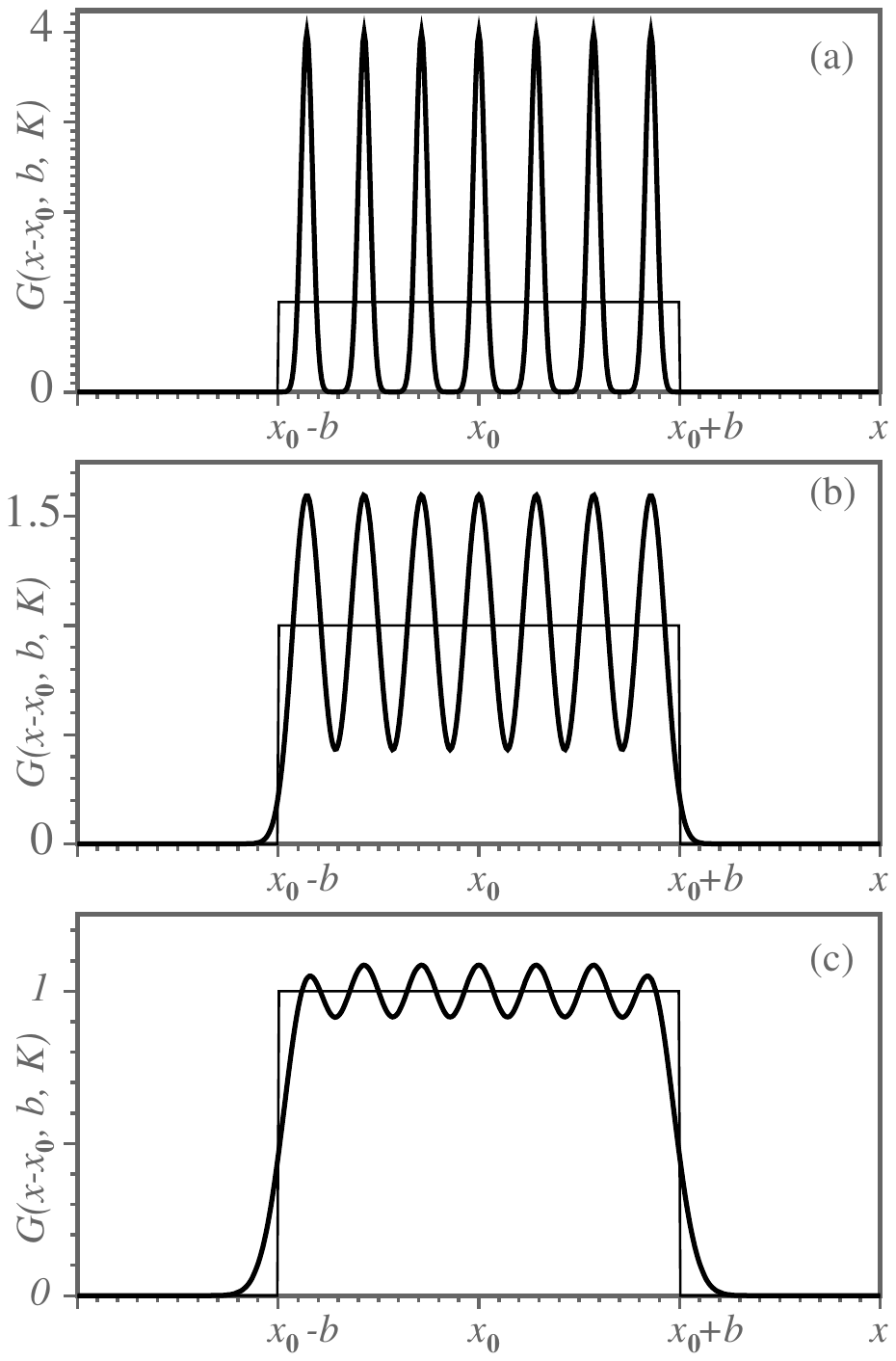}
  \end{picture}
  \caption{
  Approximation of the step function by set of the Gaussian functions presented in~(\ref{eq=38})
  with integer $K=7$ and real (a)~$\eta=0.2$; (b)~$\eta=0.5$; (c)~$\eta=0.8$.
  }
  \label{fig=18}
\end{figure}

 Let us now consider shaping the focusing spot with increasing $\eta_{1}$ but at fixed $K_{1}$.
 For definiteness, we will choose integer $K_{1}=7$
 and real $\eta_{\,1}$ ranging $0.2$, $0.5$, and $0.8$, all are smaller than~1, see Fig.~\ref{fig=18}.
 Integrals of the approximating curves are equal to a square of the step function for all cases.

\begin{figure}[htb!]
  \centering
  \begin{picture}(200,435)(29,10)
      \includegraphics[scale=0.87]{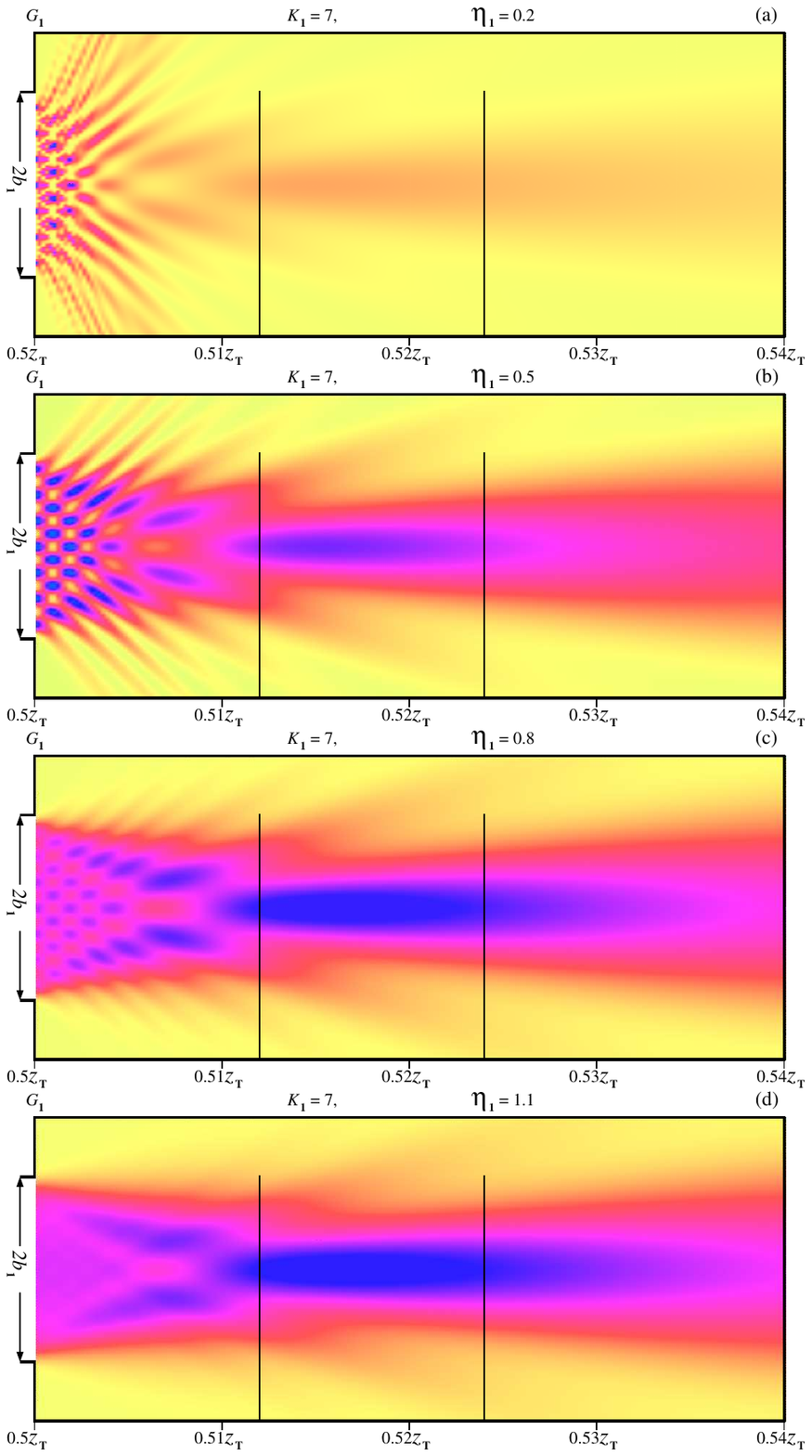}
  \end{picture}
  \caption{
  Density distribution pattern $p(x,z)$ right after the grating $G_{1}$: $z\in (0.5z_{_{\rm T}}, 0.54z_{_{\rm T}})$ and $x\in (-125, 125)$~nm:
  (a)~$\eta=0.2$; (b)~$\eta=0.5$; (c)~$\eta=0.8$; (d)~$\eta=1.1$.
  $K_{1}=7$ and $N_{0}=2$, $N_{1}=1$.
  De Broglie wavelength $\lambda_{\rm dB}=5$ pm, width of the slit $2b_{\,1}=150$~nm, and Talbot length $z_{_{\rm T}}=0.1$~m.
  }
  \label{fig=19}
\end{figure}

 It is seen that the first curve, Fig.~\ref{fig=18}(a), simulates, in fact, seven slits situated within the interval $[-b,+b]$.
 Let $2b_{1}\approx150$ nm, then a period of such a finely ruled grating is about $20$ nm.
 It corresponds to about 50 atoms situated between finely cut slits.
 The Talbot length, $z_{_{\rm T}}=0.1$~m, earlier computed represents a very large length
 as against a Talbot length, $z^{\,'}_{_{\rm T}}\sim10^{-4}$~m, for the finely ruled grating having short intervals between the finely cut slits.
 We may observe in that connection series of principal diffraction maxima from this finely cut grating.
 These maxima are partitioned by $(K_{1}-2)=5$ subsidiary maxima~\citep{Sbitnev1001}.
 All these rays are divergent in the far-field region of the finely ruled grating
 which, in turn, is situated within the near-field region of the grating $G_{1}$.

 The above described picture is shown in Fig.~\ref{fig=19}(a).
 One can see, that a central principal ray,
 diverging forward from miniature Talbot clasters packed by a triangle-like manner,
 has a narrow width before it will begin to diverge further.
 The width is positioned within $z\approx0.512z_{_{\rm T}}$ to $z\approx0.524z_{_{\rm T}}$.
 Positions of these two points in Fig.~\ref{fig=19} are marked by vertical lines.

 Let us look on the second and the third curves in Figs.~\ref{fig=18}(b) and~\ref{fig=18}(c),
 that have been simulated at $\eta_{\,1}=0.5$ and $\eta_{\,1}=0.8$, respectively.
 We see, first, because of emergence of a pedestal under the function~(\ref{eq=38})
 at increasing the parameter $\eta$, intensity of a radiation from the slits grows up considerably.
 With increasing the parameter $\eta$ the function has a trend to approximate to the step function.
 Observe that, intensity of the radiation from the slit becomes the larger the more precise the function~(\ref{eq=38})
 approximates the step function.
 At increasing the parameter $\eta$ the triangle-like pattern near the slit is seen to dissolve in the main ray.
 Together with that, the focusing spot is formed after the point $z\approx0.512z_{_{\rm T}}$.
 As the approximating function approximates to the step function, the focusing spot of the slit's beam becomes clear apparent
 within the interval from $0.512z_{_{\rm T}}$ to $0.524z_{_{\rm T}}$, see Figs.~\ref{fig=19}(b) and~\ref{fig=19}(c).
 This is an interval where rays formative the central principal maximum undergo squeezing in front of subsequent divergence~\citep{Sbitnev1001}. Fig.~\ref{fig=19}(d) shows a formed beam right away after the slit in case of the parameter $\eta>1$.

\section{\label{sec:level6}Conclusion}

 By utilizing the path integral method, we have computed the interference pattern from two gratings,
 placed in consecutive order along a particle beam.
 A wave function describing the interference is found by summing all possible trajectories of the particles passing through slits in the first grating and then in the second grating.
 This powerful method permits to describe the interference effects of the matter waves in details.

 It is instructive to attract attention to manifestation of wave-particle duality nature at describing the wave processes by this method.
 At first, we begin to consider particle paths going from a source to a detector through all possible intermediate points.
 Ensemble of all possible paths gives description of effect of propagation of a matter wave.
 Wave fronts are represented by equiphase surfaces secant the trajectory bundles.
 Observe, that these waves are those that underlie in the Huygens-Fresnel principle~\citep{Lanczos:1970}.

 The wave function found by the path integral method gives clear picture of the interference
 both between the gratings and behind the second grating.
 The most impressive observation of the interference effect is emergence of the Talbot-like carpets
 when illuminating the gratings by distributed coherent sources that are situated near the gratings, see Fig.~\ref{fig=5}(a).
 The carpets are smeared out at illumination of the gratings by incoherent matter waves, Figs.~\ref{fig=5}(b) and~\ref{fig=5}(c).
 Emergence of pedestals, supporting the interference fringes, and decreasing the visibility are conditioned by breaking coherence of the beam
 that results in smearing of the interference pattern.
 It confirms results obtained at experimental observation of the interference fringes
 fulfilled on molecular beams~\citep{Hornberger2005, JuffmannEtAl2010, GerlichEtAl2007}.

 The grating, prepared with more hard-edged slits, as was shown corrects particle flows
 in the vicinity of the slits the more powerfully than the grating with the fuzzy edged slits.
 An astonishing effect of the more hard-edged slits is that they shape focusing spots of the particle beam just after the slits,
 see the density distributions in Figs.~\ref{fig=15} and~\ref{fig=18}.
 Next, they disperse along the beam with producing ripples.
 The situation is the same as rays initially converging within a focus and diverging after.
 The focusing spot looks as a tongue-like jet passing through the beam waist.
 In particular, \citet*{Nye2002} has shown that an electromagnetic monochromatic plane wave, incident on a perfectly conducting screen
 of vanishingly small thickness that contains an infinitely long slit, reproduces behind this slit an analogous tongue-like EM jet.
 Increasing EM amplitude right after the slit is shown in this case as well.
 One may suppose, that the phenomenon of focusing by the slits having more
 hard edges is typical for many wave processes.




\end{document}